\documentclass[preprint,showpacs,superscriptaddress,aps]{revtex4-1}
\usepackage{amssymb}
\usepackage{amsmath,amssymb,graphicx}
\usepackage{graphicx}
\usepackage{dcolumn}
\usepackage{bm}
\usepackage{color}
\def\be{\begin{equation}}
\def\ee{\end{equation}}
\def\bea{\begin{eqnarray}}
\def\eea{\end{eqnarray}}

\def\lsim{\raise0.3ex\hbox{$\;<$\kern-0.75em\raise-1.1ex\hbox{$\sim\;$}}}
\def\gsim{\raise0.3ex\hbox{$\;>$\kern-0.75em\raise-1.1ex\hbox{$\sim\;$}}}

\def\slash#1{\setbox0=\hbox{$#1$}#1\hskip-\wd0\dimen0=5pt\advance
\dimen0 by-\ht0\advance\dimen0 by\dp0\lower0.5\dimen0\hbox
to\wd0{\hss\sl/\/\hss}} \hoffset=0.0cm \voffset=0.0cm

\usepackage{pstricks}
\begin{document}
\title{New physics contributions to  $\bar{B}_s \rightarrow \pi^0(\rho^0 )\,\eta^{(')} $ decays}

\author{Gaber Faisel}
\email{gaberfaisel@sdu.edu.tr}

\affiliation{{\fontsize{10}{10}\selectfont{Department of Physics,
Faculty of Arts and Sciences, S\"uleyman Demirel University,
Isparta, Turkey 32260.}}}

\affiliation{{\fontsize{10}{10}\selectfont{Department of Physics,
National Taiwan University, Taipei, Taiwan 10617.}}}

\begin{center}

\begin{abstract}
The decay modes $\bar{B}_s \rightarrow \pi^0(\rho^0 )\,\eta^{(')}
$ are dominated by electroweak penguins that are small in the
standard model. In this work we investigate the contributions to
these penguins from a model with an additional $U(1)'$gauge
symmetry and show there effects on the branching ratios of
$\bar{B}_s \rightarrow \pi^0(\rho^0 )\,\eta^{(')} $.  In a
scenario of the model, where $Z^\prime$ couplings to the
left-handed quarks vanish, we show that the maximum enhancement
occurs in the branching ratio of $\bar B^0_s\to \,\pi^0\,\eta'$
where it can reach $6$ times the SM prediction. On the other hand,
in a scenario of the model where $Z^\prime$ couplings to both
left-handed and right-handed quarks do not vanish, we find that
$Z^\prime$ contributions can enhance the branching ratio of
$B^0_s\to\,\rho^0\,\eta$ up to one order of magnitude comparing to
the SM prediction for several sets of the parameter space where
both $ \Delta M_{B_s}$ and $S_{\psi\phi}$ constraints are
satisfied.  This kind of enhancement occurs for a rather
fine-tuned point where $ \Delta M_{B_s}$ constraint on $\mid
S_{SM} (B_s) + S_{Z'} (B_s)\mid $ is fulfilled by overcompensating
the SM via $S_{Z'} (B_s) \simeq -2 S_{SM} (B_s)$.

\end{abstract}
\end{center}

\maketitle

\section{ Introduction}
The purely isospin-violating decays $\bar{B}_s \rightarrow \phi
\,\pi\,(\rho^0 ) $, $\bar{B}_s \rightarrow \pi\, \eta(\eta') $ and
$\bar{B}_s \rightarrow \rho^0\, \eta(\eta') $ are dominated by the
electroweak
penguins\cite{Fleischer:1994rs,Deshpande:1994yd,Chen:1998dta,Cheng:2009mu}.
These penguins are small in the standard model and can serve as a
probe of new physics beyond the standard model.  The decay modes
$\bar{B}_s \rightarrow \phi \,\pi\,(\rho^0 ) $  have been studied
within SM in different frameworks such as QCD factorization as in
Refs.\cite{Beneke:2003zv,Hofer:2010ee}, in PQCD  as in
Ref.\cite{Ali:2007ff} and using Soft Collinear Effective Theory
(SCET) as in Refs.\cite{Wang:2008rk,Faisel:2011kq}. The study has
been extended to include NP models namely, a modified $Z^0$
penguin, a model with an additional $U(1)'$~ gauge symmetry and
the MSSM using QCDF \cite{Hofer:2010ee}. In addition, the
investigation of NP in these decay modes has been recently
extended to include supersymmetric models with non-universal
A-term \cite{Faisel:2011kq} and two Higgs doublet models
(2HDMs)\cite{Faisel:2013nra} using SCET. The results of these
studies  showed that the additional $Z\,'$ boson of the $U(1)'$~
gauge symmetry with couplings to leptons switched off can  lead to
an enhancement in their Branching Ratios (BR) up to an order of
magnitude  making these decays  are interesting for LHCb and
future $B$ factories searches \cite{Hofer:2010ee}.

 Recently tension between the SM and data related to $b\to s\ell^+
\ell^-$ channels has become apparent. In particular LHCb has
reported deviations from the Standard Model predictions in the
processes $B\to K^*\mu^+\mu^-$ and $B_s\to \phi\mu^+\mu^-$ that
are mediated by $b\to s\mu^+ \mu^-$ transition.  Moreover the
deviations include the process $B\to K\mu^+\mu^-$ through the
ratio $R_K$ defined as

\be R_K = \frac{Br (B\to K\mu^+\mu^-)}{Br(B\to K e^+e^-)} \ee

These anomalies can be naturally accommodated in $Z'$ models as
have been found in
Refs.\cite{Descotes-Genon:2013wba,Gauld:2013qba,Buras:2013qja,Gauld:2013qja,Buras:2013dea,Altmannshofer:2014cfa}.
These findings serve as a general phenomenological motivation for
$Z'$ models and for the search of analogous tensions in hadronic
$B$ decays. In this work we investigate the phenomenological
implications of a leptophobic $Z'$ model on the decay modes
$\bar{B}_s \rightarrow \pi\, \eta(\eta') $ and $\bar{B}_s
\rightarrow \rho^0\, \eta(\eta') $. In this model $Z'$ couplings
to quarks are not related to their couplings to leptons and thus
can avoid the tight constraints from semileptonic decays
\cite{Hofer:2010ee}.

 The decay modes  $\bar{B}_s \rightarrow \pi\,
\eta(\eta') $ and $\bar{B}_s \rightarrow \rho^0\, \eta(\eta') $
have been studied within SM using different frameworks such as
Naive Factorization (NF)\cite{Deshpande:1994yd}, generalized
factorization \cite{Chen:1998dta}, POCD
\cite{Xiao:2006gf,Ali:2007ff}  and QCDF
\cite{Sun:2002rn,Cheng:2009mu}. On the other hand, using SCET, an
investigation of $\bar{B}_s \rightarrow \pi\, \eta(\eta') $ has
been carried out in Ref.\cite{Williamson:2006hb} while  the decay
modes $\bar{B}_s \rightarrow \rho^0\, \eta(\eta') $ has been
studied in Ref.\cite{Wang:2008rk}.  NP effects namely 2HDMs has
been investigated  in these decay modes in Ref.
\cite{Deshpande:1994yd} using NF and using generalized
factorization in Ref. \cite{Zhang:2000ic}.   In our study we will
adopt SCET as a framework for the calculation of the
amplitudes\cite{Bauer:2000ew,Bauer:2000yr,Chay:2003zp,Chay:2003ju}.

  SCET provides a systematic and  rigorous way to deals with the
processes in which energetic quarks and gluons have different
momenta modes such as hard, soft and collinear modes. The power
counting in SCET  reduces the complexity of the calculations.  In
addition, the factorization formula given by SCET is perturbative
to all powers in $\alpha_s$ expansion.

 This paper is organized as follows. In Sec.~\ref{sec:formalism},
we  review the decay amplitude for $B \to M_1M_2$ within SCET
framework. Accordingly, we present the SM predictions of the
branching ratios of the decay modes under the study in
Sec.~\ref{SMse}. Then we proceed to analyze NP contributions
namely the non universal $Z^{\prime}$ model in section
~\ref{NPCon}. Finally, we give our conclusion in
Sec.~\ref{sec:conclusion}.

\section{ $\bar{B}_s \rightarrow \pi^0(\rho^0 )\,\eta^{(')} $ decays in SCET \label{sec:formalism} }

\begin{table}
\begin{center}
\begin{tabular}{|c|c|c|c|c|}
  \hline
  Decay mode & $T_1$ & $T_2$ & $T_{1g}$ & $T_{2g}$ \\
  \hline
  $\bar{B}_s \rightarrow \eta_s\,\pi^0 $ & 0 & $\frac{1}{\sqrt{2}} (c^s_2-c^s_3)$ & 0 &$\frac{1}{\sqrt{2}} (c^s_2-c^s_3)$  \\
  $\bar{B}_s \rightarrow \eta_s\, \rho^0  $ & 0 & $\frac{1}{\sqrt{2}} (c^s_2+c^s_3)$ & 0 & $\frac{1}{\sqrt{2}} (c^s_2+c^s_3)$ \\
  $\bar{B}_s \rightarrow \eta_q\,\pi^0\ $ & 0 & 0 & 0 & $(c^s_2-c^s_3)$  \\
  $\bar{B}_s \rightarrow  \eta_q\,\rho^0$ & 0 & 0 & 0 & $(c^s_2+c^s_3)$ \\
  \hline
\end{tabular}
\end{center}
\caption{ Hard kernels of $\bar{B}^0_s\to \eta_{s,q}\, \pi^0
(\rho^0)$ decays. The hard kernels $T_{iJ}$, $T_{iJg}$, for
$i=1,2$, can be obtained through the replacement $ c_i^s\to b_i^s
$}\label{Hardk}
\end{table}

The decay amplitude of $B$ meson into two light final states
mesons $ M_1 , M_2$  at LO in $\alpha_s(m_b)$ expansion in SCET
can be written as \cite{Wang:2008rk}:
\begin{eqnarray}
  A(B\to M_1 M_2)&=&\frac{G_F}{\sqrt{2}} m_{{B}}^2 \left\{
  f_{M_1}\left[ \zeta_J^{{B} M_2}\int\negthickspace du\phi_{M_1}(u)  T_{1J}(u)
  +\zeta_{Jg}^{ {B} M_2}\int\negthickspace du\phi_{M_1}(u)   T_{1Jg}(u)\right] \right.   \nonumber\\
  &&\left.+ f_{M_1}  (T_{1}
  \zeta^{{B} M_2}+T_{1g}\zeta_g^{{B} M_2})
  +A^{M_1 M_2}_{cc}+ (1\leftrightarrow 2)\right\},\label{ampM1M2}
\end{eqnarray}

here $ M_1 M_2$ can be $PP$ or $PV$ where $P$ stands for
pseudoscalar meson and $V$ stands for vector meson and
$\phi_{M}(u)$ is the light-cone distribution amplitude (LCDA) of
the meson $M$. $A^{M_1 M_2}_{cc}$ represents the non-perturbative
long distance charm contribution to the amplitude. The hadronic
parameters $\zeta^{{B} M}$, $\zeta^{ {B} M}_g$, $\zeta_J^{{B} M}$
and $\zeta_{Jg}^{{B} M}$, in the framework of SCET,  are treated
as non-perturbative parameters that can be fitted using the
experimental data of the branching fractions and CP asymmetries of
the  non leptonic $B$ and $B_s$ decays \cite{Bauer:2005kd,
Williamson:2006hb,Jain:2007dy,Wang:2008rk}. The hard kernels
$T_{i}$, $T_{ig}$, $T_{iJ}(u)$ and $T_{iJg}(u)$ for $i=1,2$ are
functions of the Wilson coefficients of the weak effective
Hamiltonian. The expressions of these kernels  for a certain $B\to
M_1 M_2$ decay mode in the case of SM can be obtained using the
formulas given in the appendix of Ref.\cite{Wang:2008rk}.

 In  many extensions  of the SM the weak effective Hamiltonian can have
new set of operators $\tilde Q_i$ that are obtained by flipping
the chirality of the SM four-quark operators from left to right.
Following a similar treatment to that in Ref.\cite{Wang:2008rk} we
find that the effect of these new operators can be incorporated in
the expressions of the hard kernels $T_{i}$, $T_{ig}$, $T_{iJ}(u)$
and $T_{iJg}(u)$. In Table (\ref{Hardk}) we  present the explicit
expressions of the hard kernels $T_{i}$, $T_{ig}$, $T_{iJ}(u)$
relevant to the decay channels $\bar{B}_s\to \eta_s\, \pi^0$,
$\bar{B}_s\to \eta_{q}\, \pi^0$, $\bar{B}_s\to \eta_{s}\, \rho^0$
and $\bar{B}_s\to \eta_{q}\, \rho^0$. The Coefficients $c^s_i$ and
$b^s_i$ are functions of Wilson coefficients of the weak effective
Hamiltonian. After extending the SM weak effective Hamiltonian to
include right-handed operators $\tilde Q_i$ generated by NP we
find that
\begin{eqnarray}
c_{2}^{(s)}&=&\lambda_u^{(s)}\Big[C_{2}-\tilde{C}_2+\frac{1}{N_c}(C_{1}-\tilde{C}_1)\Big]
-\frac{3}{2}
\lambda_t^{(s)}\Big[C_{9}-\tilde{C}_9+\frac{1}{N_c}(C_{10}-\tilde{C}_{10})
\Big],\nonumber\\
c_3^{(s)}&=&
-\frac{3}{2}\lambda_t^{(s)}\Big[C_7-\tilde{C}_7+\frac{1}{N_c}(C_8-\tilde{C}_8)\Big],\eea

and

\bea b_{2}^{(s)}
&=&\lambda_u^{(s)}\Big[C_{2}+\tilde{C}_2+\frac{1}{N_c}\Big(1-\frac{m_b}{\omega_3}\Big)(C_{1}+\tilde{C}_1)\Big]
-\frac{3}{2}\lambda_t^{(s)}\Big[C_{9}+\tilde{C}_9+\frac{1}{N_c}\Big(1-\frac{m_b}{\omega_3}\Big)(C_{10}+\tilde{C}_{10})\Big],
\nonumber\\
b_3^{(s)} &=&- \frac{3}{2} \lambda_t^{(s)}\Big[C_7-\tilde{C}_7+
\frac{1}{N_c}
\Big(1-\frac{m_b}{\omega_2}\Big)(C_8-\tilde{C}_8)\Big]
,\label{eq:wcsubleadingpower}
\end{eqnarray}
where ${C}_i = C^{SM}_i + C^{NP}_i $ and $\tilde{C}_i = \tilde
C^{SM}_i +\tilde C^{NP}_i $. The Wilson coefficients $\tilde{C}_i$
correspond to the four-quark operators in the weak effective
Hamiltonian that have right chirality. In the SM such operators
are absent and hence $\tilde{C}^{SM}_i=0$. In
Eq.(\ref{eq:wcsubleadingpower}) we have $\omega_2=u m_{\bar{B}_s}
$ and $\omega_3=-\bar u m_{\bar{B}_s}$ with $u$ is the momentum
fraction of the positive quark in the emitted meson and $N_c=3$.
From charge conjugation and isospin we have   $\phi_{\pi
(\rho)}(u)= \phi_{\pi (\rho)}(1-u)$ \cite{Jain:2007dy}.  Thus we
can write
 \be \int^1_0 du \frac{\phi_{M}(u)}{\bar u} = \int^1_0 du
\frac{\phi_{M}(1-u)}{1-u} = \int^1_0 du \frac{\phi_{M}(u)}{u}
=\langle\chi^{-1}\rangle _{M} \ee  for $M=\pi$ and $M=\rho$. This
relation together with the relation

\be \int^1_0  du \phi_{M}(u)=1\ee allow us to perform the
integrals in eq.(\ref{ampM1M2}) and express the results in terms
of the hadronic parameter $\langle\chi^{-1}\rangle _{M}$. For the
physical states  $\eta$ and $\eta^{\prime}$ they are related to
the flavor basis $\eta_s$  and $\eta_q $ through
\cite{Williamson:2006hb}:

\begin{equation}
\left(
\begin{array}{c}
\eta  \\
\eta^{\prime}
\end{array}
\right) = \left(
\begin{array}{cc}
\cos \phi & - \sin \phi\\
\sin \phi & \cos \phi
\end{array}
\right)~ \left(
\begin{array}{c}
 \eta_q  \\
 \eta_s
\end{array}
\right).
\end{equation}
Where the mixing angle is measured as $\phi= 46^\circ$
\cite{Michael:2013gka}. Upon using this relation we can easily
calculate the decay amplitudes of $\bar{B}_s \to \eta M$ and
$\bar{B}_s \to \eta^{\prime} M$ for $M=\pi$ and $M=\rho$.

\section{ $\bar{B}_s \rightarrow \pi^0(\rho^0 )\,\eta^{(')} $ decays in the standard
model \label{SMse} }

In this section we give our predictions for the branching ratios
of $\bar{B}_s \rightarrow \pi^0(\rho^0 )\,\eta^{(')} $ decays in
the standard model. In our analysis, we use the different set of
values given in Refs.\cite{Wang:2008rk} for the hadronic
parameters $\zeta^{{B} M}$, $\zeta^{ {B} M}_g$, $\zeta_J^{{B} M}$
and $\zeta_{Jg}^{{B} M}$ corresponding to the two solutions
obtained from the $\chi^2$ fit and assuming a $20\%~$ error in
their values due to the SU(3) symmetry breaking. It should be
noted that the values that enter the SCET predictions are obtained
from a fit to data assuming SM Wilson coefficients. They are valid
for NP analysis provided the NP contribution to those channels
which dominate the fit is small compared to the SM contribution.
This is the case for the considered $Z'$ scenarios. We use for the
inverse moment of the $\rho$ meson light-cone distribution
amplitude $\langle\chi^{-1}\rangle _{\rho}= 3.45$
\cite{Ball:2007rt} and $\langle\chi^{-1}\rangle _{\pi}= 2.9\pm
0.4$\cite{Bakulev:2003cs}.

The amplitudes of $\bar{B}_s \rightarrow \pi^0(\rho^0
)\,\eta^{(')} $ decays in the SM can be obtained  by setting  $
\tilde C^{SM}_i = C^{Z'}_i = \tilde C^{Z'}_i = 0 $. For $\bar{B}_s
\rightarrow \pi^0\,\eta^{(')} $ decays we obtain

\bea {\cal A}_1(\bar{B}^0_s\to \eta\, \pi^0)\times 10^6&\simeq& -
5.6\, C^{SM}_9\,\lambda^s_{c} -(2.3\, C^{SM}_1 + 3.7\,
C^{SM}_2 )\lambda^s_{u}\nonumber\\
{\cal A}_2(\bar{B}^0_s\to \eta\, \pi^0)\times 10^6&\simeq&  -
5.1\, C^{SM}_9\,\lambda^s_{c} - (1.6\, C^{SM}_1 + 3.4\,
C^{SM}_2)\lambda^s_{u}\nonumber\\{\cal A}_1(\bar{B}^0_s\to \eta'
\, \pi^0)\times 10^6&\simeq& 0.4\, C^{SM}_9\,\lambda^s_{c}
+ (0.1\, C^{SM}_1+0.3\, C^{SM}_2)\lambda^s_{u}\nonumber\\
{\cal A}_2(\bar{B}^0_s\to \eta' \, \pi^0)\times 10^6&\simeq& 1.8\,
C^{SM}_9\,\lambda^s_{c} + (2.7\, C^{SM}_1+1.2\, C^{SM}_2
)\lambda^s_{u}\label{smp}\eea

while for $\bar{B}_s \rightarrow \rho^0 \,\eta^{(')} $ decays we
obtain

 \bea {\cal A}_1(\bar{B}_s\to \eta\,\rho^0 )\times
10^6&\simeq& - 6.3\, C^{SM}_9 \,\lambda^s_{c} - (3.4\, C^{SM}_1
+4.2\,C^{SM}_2 )\lambda^s_{u} \nonumber\\
{\cal A}_2(\bar{B}_s\to \eta\,\rho^0 )\times 10^6&\simeq& - 3.0\,
C^{SM}_9 \,\lambda^s_{c} - (1.6\, C^{SM}_1+2.0\, C^{SM}_2
)\lambda^s_{u}\nonumber\\{\cal A}_1(\bar{B}_s\to \eta'\,\rho^0
)\times 10^6&\simeq& 3.3\,C^{SM}_9 \,\lambda^s_{c} + (0.8\,
C^{SM}_1+2.2\, C^{SM}_2 )\lambda^s_{u}\nonumber\\{\cal
A}_2(\bar{B}_s\to \eta'\,\rho^0 )\times 10^6&\simeq& 8.2\,
C^{SM}_9\,\lambda^s_{c} + (6.3\, C^{SM}_1+5.4\, C^{SM}_2
)\lambda^s_{u} \label{smrho1}\eea

where we have used the unitarity of the CKM matrix to write
$\lambda^s_{t}= - \lambda^s_{u}-\lambda^s_{c}$ and also the
hierarchy of the SM Wilson coefficients $C^{SM}_1\gg C^{SM}_{i}$
for $i=2,3,4,5,6,7,8,9,10$ and $C^{SM}_9\gg C^{SM}_{i}$ for
$i=7,8,10$. The amplitudes ${\cal A}_1$ and ${\cal A}_2$ refers to
solutions $1$ and $2$ of the SCET parameters respectively. From
the CKM matrix we can write to a good approximation $\lambda^s_{c}
\simeq Re (\lambda^s_{c}) \simeq 0.04$ and $|\lambda^s_{u}| \simeq
2\times 10^{-2} \lambda^s_{c}$. At leading order we have

\be C^{SM}_1= 1.1,\,\,\,\,\, C^{SM}_2= -0.253,\,\,\,\,\, C^{SM}_9=
-10.3\times 10^{-3}\ee

 Clearly the real parts of the amplitudes in  Eqs(\ref{smp},\ref{smrho1})
 are dominant by the terms proportional to $C^{SM}_9\,\lambda^s_{c}
\simeq -4\times 10^{-4}$. This
 can be attributed to several reasons. First, the cancellation
 that take places in the $\lambda^s_{u}$ terms of the amplitudes
 due to the sign difference between $C^{SM}_1$ and $C^{SM}_2$. Second, the sign
 difference between the terms proportional to $\lambda^s_{c}$ and
 $\lambda^s_{u}$ after taking into account the minus sign
 of the Wilson coefficient $C^{SM}_9$. And finally due to the hierarchy
$|\lambda^s_{u}| \simeq 2\times 10^{-2} \lambda^s_{c}$. Another
remark, the imaginary parts of the amplitudes in
Eqs(\ref{smp},\ref{smrho1}) are suppressed as they are
proportional to $|\lambda^s_{u}| \simeq 2\times 10^{-2}
\lambda^s_{c} \simeq 10 ^{-4}$. As a consequence the predicted
branching ratios for these decay modes are small as shown in Table
(\ref{branch}). The last two columns give the predictions
corresponding to the two solutions of the SCET parameters obtained
from the $\chi^2$ fit. The errors on the SCET predictions are due
to SU(3) breaking effects and errors due to SCET parameters
respectively.

\begin{table}
\begin{center}
\begin{tabular}{|c|c|c|c|c|c|c|}
  \hline
  Decay channel & QCDF & PQCD & SCET solution $1$ & SCET solution $2$\\
  \hline
  $\bar{B}_s\to \eta\, \pi^0$  & $0.075_{-0.012-0.025-0.010-0.007}^{+0.013+0.030+0.008+0.010}$ & $ 0.05 _{-0.02-0.01-0.00}^{+0.02+0.01+0.00}$&
 $0.037 _{-0.010-0.006}^{+0.010+0.006} $& $0.031_{-0.009-0.003}^{+0.009+0.003}$
   \\
  $\bar{B}_s\to \eta'\, \pi^0$  & $ 0.11_{-0.02-0.04-0.01-0.01}^{+0.02+0.04+0.01+0.01}$ & $0.11_{-0.03-0.01-0.00}^{+0.05+0.02+0.00}$
  & $0.0002_{-0.001-0.001}^{+0.001+0.001}$ & $0.033_{-0.010-0.010}^{+0.010+0.010}$ \\
  $\bar{B}_s\to \eta\, \rho^0$  & $0.17_{-0.03-0.06-0.02-0.01}^{+0.03+0.07+0.02+0.02}$ & $ 0.06_{-0.02-0.01-0.00}^{+0.03+0.01+0.00}$&
  $ 0.055_{-0.018-0.017}^{+0.018+0.017} $& $ 0.012_{-0.009-0.004}^{+0.009+0.004}$ \\
  $\bar{B}_s\to \eta'\, \rho^0$  & $ 0.25_{-0.05-0.08-0.02-0.02}^{+0.06+0.10+0.02+0.02}$ & $0.13_{-0.04-0.02-0.01}^{+0.06+0.02+0.00}$
  & $0.013_{-0.009-0.016}^{+0.009+0.016}$ & $0.148_{-0.045-0.043}^{+0.045+0.043}$ \\
  \hline
\end{tabular}
 \end{center}
\caption{ Branching ratios of $\bar{B}^0_s\to \eta^{(')}\, \pi^0$
and $\bar{B}_s\to \eta^{(')}\, \rho^0$ decays in $10^{-6}$ units.
The last two columns  give the predictions corresponding to the
two solutions of the SCET parameters obtained from the $\chi^2$
fit. On the SCET predictions the errors are due to SU(3) breaking
effects and errors due to SCET parameters respectively. For a
comparison with previous studies in the literature, we list the
results evaluated in QCDF \cite{Beneke:2003zv}, PQCD
\cite{Ali:2007ff}.}\label{branch}
\end{table}

\section{ $Z'$ model contributions to $\bar{B}_s \rightarrow \pi^0(\rho^0 )\,
\eta^{(')}$ decays\label{NPCon}}

One of the possible extension of the SM is to enlarge the SM gauge
group to include additional $U(1)^{\prime}$ gauge group. This
possibility is well-motivated in several beyond SM theories such
as theories with large extra dimensions\cite{Masip:1999mk} and
grand unified theories\cite{E6}. As a consequence of the
$U(1)^{\prime}$ gauge symmetry a new gauge boson, $Z^{\prime}$,
arises. Basically $Z^{\prime}$ can have either family universal
couplings or family non-universal couplings to the SM fermions. In
the case that $Z^{\prime}$ gauge couplings are family universal
they remain diagonal even in the presence of fermion flavor mixing
by the GIM mechanism\cite{Langacker:2000ju}. On the other hand and
in some models like string models it is possible to have
family-non universal $Z^{\prime}$ couplings, due to the different
constructions of the different
families\cite{Chaudhuri:1994cd,Cleaver:1997jb,Cleaver:1998gc,
Langacker:2000ju}. This scenario with family-non universal
couplings has theoretical and phenomenological motivations. For
instance, possible anomalies in the Z -pole $b \bar{b}$
asymmetries suggest that the data are better fitted with a
non-universal $Z^{\prime}$ \cite{Chang:2013hba}. Recent studies
about the phenomenology of $Z^{\prime}$ has been performed in
Refs.\cite{Altmannshofer:2009ma,Buras:2012dp,Buras:2012jb,Buras:2013qja,Buras:2013rqa,Buras:2013td,Buras:2013uqa,Buras:2014fpa}.
For a detailed review about the physics of $Z^{\prime}$
gauge-bosons we refer to Ref.\cite{Langacker:2008yv}.

 In our analysis we will follow
Refs.~\cite{Langacker1,Langacker2,chiang1,Hofer:2010ee,Grossman:1999av,Barger:2009eq,Chang:2009wt,Barger:2009qs,Chang:2013hba,Altmannshofer:2009ma,Buras:2012dp,Buras:2012jb,Buras:2013qja,Buras:2013rqa,Buras:2013td,Buras:2013uqa,Buras:2014fpa}
and consider a non-universal $Z^{\prime}$ couplings in a way
independent to a specific $Z^{\prime}$ model. Neglecting $Z$-$Z'$
mixing and assuming the absence of exotic fermions that can mix
with the SM fermions through the $Z'$ couplings, the
quark-antiquark-$Z^{\prime}$ interaction Lagrangian can be written
as \cite{Grossman:1999av,Barger:2009qs,Hofer:2010ee}
\be\label{c20} {\cal L}^{eff}_{Z^{\prime}} =
-\frac{g_{U(1)'}}{2\sqrt{2}}\sum_{ij} \bar{q}_i
\left[\zeta_L^{ij}\gamma^{\mu}(1-\gamma_5)
+\zeta_R^{ij}\gamma^{\mu}(1+\gamma_5)\right]q_j Z^{\prime}_{\mu}.
\ee where $i$ and $j$  denote different quark flavours of the same
type quarks. In order to simplify our analysis we introduce the
parameters \be i \Delta ^{ij}_{L,R} \equiv
-\frac{g_{U(1)'}}{\sqrt{2}}\zeta_{L,R}^{ij} \ee In terms of these
parameters we find that $Z^{\prime}$ contributions to the
electroweak penguins relevant to our decay processes, at the
electroweak scale, are given as

\begin{eqnarray}\label{c21}
C^{Z^{\prime}}_{7}\, =\,\,\frac{4}{3} \frac{ M_W^2
\Delta_L^{sb}}{g^2 M_{Z'}^2\lambda_t^{(s)}}\, \left(\Delta_R^{uu}
-\Delta_R^{dd} \right)\,, \hspace{1.1cm}\qquad &&
\tilde{C}^{Z^{\prime}}_7\, =\,\,\frac{4}{3} \frac{ M_W^2
\Delta_R^{sb}}{g^2 M_{Z'}^2\lambda_t^{(s)}}\, \left(\Delta_L^{uu}
-\Delta_L^{dd} \right)\,, \nonumber\\
C^{Z^{\prime}}_{9}\, =\, \,\frac{4}{3}\frac{ M_W^2
\Delta_L^{sb}}{g^2 M_{Z'}^2\lambda_t^{(s)}}\,
\left(\Delta_L^{uu}-\Delta_L^{dd}\right)\,, \hspace{2.1cm} &&
\tilde{C}^{Z^{\prime}}_9\, =\, \,\frac{4}{3}\frac{ M_W^2
\Delta_R^{sb}}{g^2 M_{Z'}^2\lambda_t^{(s)}}\,
\left(\Delta_R^{uu}-\Delta_R^{dd}\right)\,.
\end{eqnarray}
The $SU(2)_L$ invariance implies that
$\zeta_{L}^{uu}=\zeta_{L}^{dd}$ \cite{Hofer:2010ee}. As a
consequence $\Delta_{L}^{uu}=\Delta_{L}^{dd}$ and thus we are left
with only two non-vanishing coefficients
\begin{eqnarray}\label{c22}
C^{Z^{\prime}}_{7}\, &=&\,\,\frac{4}{3} \frac{ M_W^2
\Delta_L^{sb}}{g^2 M_{Z'}^2\lambda_t^{(s)}}\, \left(\Delta_R^{uu}
-\Delta_R^{dd} \right)\,, \nonumber\\
\tilde{C}^{Z^{\prime}}_9\, &=&\, \,\frac{4}{3}\frac{ M_W^2
\Delta_R^{sb}}{g^2 M_{Z'}^2\lambda_t^{(s)}}\,
\left(\Delta_R^{uu}-\Delta_R^{dd}\right)\,.\label{ZWilson}
\end{eqnarray}

Last equation indicates that $Z'$ contributions to the Electroweak
Wilson coefficients vanish in the case of $\Delta ^{uu}_{R} =
\Delta ^{dd}_{R} = 0$ or $\Delta ^{uu}_{R} = \Delta ^{dd}_{R}$.

We discuss now the constraints imposed on the $\Delta ^{ij}_{L,R}$
parameters. To avoid tight constraints from semileptonic decays we
consider $Z'$ model with vanishing couplings to leptons. In this
model $Z'$ mass is much less constrained \cite{Hofer:2010ee}. This
can be explained as leptophobic $Z'$ bosons can avoid detection
via traditional Drell-Yan processes. This choice can be adopted as
the couplings of the $Z'$ boson to quarks are not related to their
couplings to leptons. This leptophobic$Z'$ boson can appear in
models with an $E_6$ gauge symmetry \cite{Rizzo:1998ut}.

 The most stringent constraints on the couplings $\Delta_L^{sb}$
and $\Delta_R^{sb}$ stem from $B_s-\overline{B}_s$ mixing. The
effective Hamiltonian governs $B_s-\overline{B}_s$ mixing can be
written as \cite{Buras:2001ra,Buras:2012fs,Buras:2012jb}

\be{\cal H}^{\Delta  f=2}_{eff}= C^{VLL}_1 Q^{VLL}_1+ C^{VRR}_1
Q^{VRR}_1+C^{LR}_1 Q^{LR}_1+ C^{LR}_2 Q^{LR}_2\ee where the
four-quark operators are given as \bea Q^{VLL}_1 &=& \left[ \bar
b_\alpha \gamma^\mu P_L s_\alpha \right]\left[
\bar b_\beta \gamma^\mu P_L s_\beta \right] \,, \nonumber\\
Q^{VRR}_1 &=& \left[ \bar b_\alpha \gamma^\mu P_R s_\alpha
\right]\left[\bar b_\beta \gamma^\mu P_R s_\beta \right] \,, \nonumber\\
Q^{LR}_1 &=& \left[ \bar b_\alpha \gamma^\mu P_L s_\alpha
\right]\left[
\bar b_\beta \gamma^\mu P_R s_\beta \right] \,, \nonumber\\
Q^{LR}_2 &=& \left[ \bar b_\alpha  P_L s_\alpha \right]\left[ \bar
b_\beta P_R s_\beta \right] \,,\label{BsBs} \eea

The $\Delta B_s = 2$ mass difference is given as
\cite{Buras:2012jb} \be \Delta M_{B_s}=\frac{G^2_F}{6\pi^2}M^2_W
m_{B_s}|\lambda^s_t|^2 F^2_{B_s}\hat{B}_{B_s}\eta_B |S(B_s)|\ee
The expression of $S(B_s)$ can be expressed as \cite{Buras:2012jb}
\be S(B_s)=S_0(x_t)+[\Delta S(B_s)]_{VLL}+[\Delta
S(B_s)]_{VRR}+[\Delta S(B_s)]_{LR} \equiv |S(B_s)| e^{i
\theta^{B_s}_S}\label{SBS}\ee where the loop function $S_0(x_t)$
stems from the SM contribution to the $\Delta B_s = 2$ mass
difference

\be S_0(x_t)=\frac{4 x_t-11 x^2_t+x^3_t}{4(1-x_t)^2}-\frac{3
x^2_t\log x_t}{2(1-x_t)^3}\ee with $x_t=  \frac{m^2_t}{m^2_W}$.
The rest of quantities in $S(B_s)$ account for $Z^{\prime}$
contribution to the $\Delta B_s = 2$ mass difference. The
expressions for $[\Delta S(B_s)]_{VLL(VRR)}$ are given as
\cite{Buras:2012jb}

\bea [\Delta
S(B_s)]_{VLL(VRR)}&=&\big[\frac{\Delta_{L(R)}^{bs}}{\lambda^s_t}\big]^2\frac{4\tilde{r}}
{M_{Z^\prime}^2 g^2_{SM}}\label{VLVR}\eea

Here $\tilde{r}$ is a factor that accounts for QCD renormalization
group effects. Explicit expressions for $\tilde{r}$ and $g^2_{SM}$
can be found in Ref.\cite{Buras:2012dp}. Turning now to the
expression of $[\Delta S(B_s)]_{LR}$ one finds that
\cite{Buras:2012jb}
 \bea [\Delta S(B_s)]_{LR}&=& \frac{\Delta_L^{bs}
\Delta_R^{bs}}{M_{Z^\prime}^2
T(B_s)}\big[C^{LR}_1(\mu_{Z'})\langle
Q^{LR}_1(\mu_{Z'},B_s)\rangle+C^{LR}_2(\mu_{Z'})\langle
Q^{LR}_2(\mu_{Z'},B_s)\rangle\big]\label{VLR}\eea

 where \bea
T(B_s)&=& \frac{G^2_F}{12\pi^2}M^2_W
m_{B_s}|\lambda^s_t|^2 F^2_{B_s}\hat{B}_{B_s}\eta_B\nonumber\\
C^{LR}_1(\mu_{Z'})&=&1+\frac{\alpha_s}{4\pi}\big(-\log\frac{M_{Z^\prime}^2}
{\mu^2_{Z'}}-\frac{1}{6}\big)\,,\nonumber\\
C^{LR}_2(\mu_{Z'})&=&\frac{\alpha_s}{4\pi}\big(-6\log\frac{M_{Z^\prime}^2}
{\mu^2_{Z'}}-1\big)\,.\eea

The central values of the matrix elements $\langle
Q^{LR}_{1,2}(\mu_{Z'},B_s)\rangle$ can be found in Table 1 in
Ref.\cite{Buras:2012jb}. In order to take the experimental and the
hadronic uncertainties into account we follow
Ref.\cite{Buras:2012jb} and require that the theory to reproduce
the data for $\Delta M_{B_s}$ within $\pm 5\%$. Thus for $\Delta
M^{Exp}_{B_s}=17.761(22)\, ps^{-1}$ \cite{Amhis:2012bh} the
allowed range reads

\be 16.9/ ps \leq \Delta M_{B_s} \leq 18.6 /ps \label{mBs}\ee

The previous relation can be used to set constraints on the
parameters $\Delta_{L,R}^{bs}$ once the values of $M_{Z'}$ are
given. The most stringent constraints on $M_{Z'}$ are provided by
CMS experiment \cite{Chatrchyan:2012oaa}. For the sequential $Z'$
model the lower bound for $M_{Z'}$ is 2.59 TeV while in other
models values as low as 1 TeV are still possible.

In addition to the constraint from $\Delta M_{B_s}$ we need to
take into account the constraint from $S_{\psi\phi}$ which can be
defined as \cite{Buras:2012jb}

 \be S_{\psi\phi}=
\sin(2|\beta_s|-2\phi_{B_s})\ee where the phases $\beta_s$ and
$\phi_{B_s}$ are defined by \be V_{ts}= -
|V_{ts}|e^{-i\beta_s},\,\, \,\,\,\,\,\,\,\,\,2\phi_{B_s} =
-\theta^{B_s}_S \ee

with $\beta_s \simeq -1^\circ$ and $\theta^{B_s}_S$ is the phase
of $S(B_s)$ given in Eq.(\ref{SBS}). The LHCb measurement of
$S_{\psi\phi}$ reads \cite{Clarke:2012hhi}
 \be S_{\psi\phi}= 0.002\pm 0.087\ee

To use $S_{\psi\phi}$ as a constraint on the parameter space, we
follow Ref.\cite{Buras:2012jb} and require $S_{\psi\phi}$ to vary
in the range

\be - 0.18 \leq S_{\psi\phi} \leq 0.18 \label{Sphi}\ee

In our analysis we will consider two scenarios. In the first
scenario, the Right-Handed Scenario (RHS), we assume
$\Delta_{R}^{ij} \neq 0$ and $\Delta_{L}^{ij}= 0$. Note that the
scenario with $\Delta_{L}^{ij} \neq 0$ and $\Delta_{R}^{ij}= 0$ is
not interesting for our decay modes as this scenario leads to the
vanishing of the Wilson coefficients. In the second scenario, the
Left-Right Scenario (LRS), we assume $\Delta_{R}^{ij} \neq 0$ and
$\Delta_{L}^{ij} \neq 0$. In Ref.\cite{Buras:2012jb} a scenario
with a left-right symmetry in the $Z'$-couplings to quarks i.e.
$\Delta_{L}^{ij}= \Delta_{R}^{ij}$ has been adopted.  This
scenario is not relevant to our decay modes as it leads to
vanishing $Z'$ contributions to the amplitudes. This can be
explained as the $SU(2)_L$ invariance implies that
$\Delta_{L}^{uu}= \Delta_{L}^{dd}$ and  hence $\Delta_{R}^{uu}-
\Delta_{R}^{dd} = \Delta_{L}^{uu}- \Delta_{L}^{dd} = 0$ leading to
vanishing $C^{Z'}_7$ and $\tilde C^{Z'}_9$. Thus in our LRS
scenario we take $\Delta_{L}^{ij} \neq \Delta_{R}^{ij}$.

\begin{figure}[tbhp]
\includegraphics[width=5cm,height=5cm]{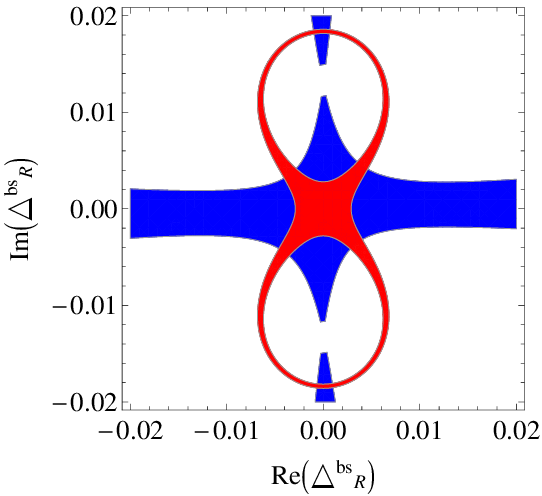}
\hspace{0.2cm}
\includegraphics*[width=5cm,height=5cm]{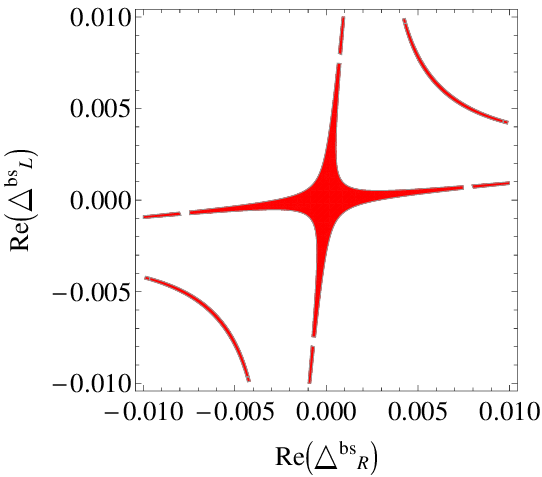}
\hspace{0.2 cm}
\includegraphics*[width=5cm,height=5cm]{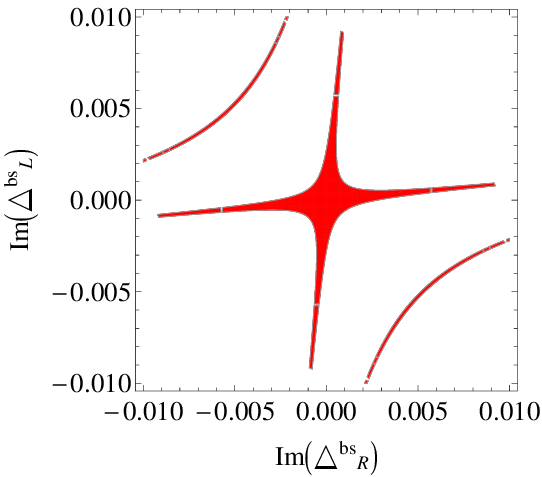}
\medskip
\caption{ Left: allowed regions in the $Re (\Delta_{R}^{bs}) - Im
(\Delta_{R}^{bs})$ plane in RHS where red (blue) color corresponds
to the bounds on $ \Delta M_{B_s}$ ($S_{\psi\phi}$). Middle:
allowed regions in $  Re (\Delta_{R}^{bs})- Re (\Delta_{L}^{bs})$
plane in LRS from the bounds on $ \Delta M_{B_s}$ corresponding to
the case $Im (\Delta_{L}^{bs})=Im (\Delta_{R}^{bs})=0$. Right:
allowed regions in the in $ Im (\Delta_{R}^{bs})- Im
(\Delta_{L}^{bs})$ plane in LRS from the bounds on $ \Delta
M_{B_s}$ corresponding to the case $Re(\Delta_{L}^{bs})=Re
(\Delta_{R}^{bs})=0$. In all plots we take $M_{Z^\prime}=1$ TeV}
\label{ZBs}
\end{figure}

 We start our analysis by investigating the parameter space in
the two scenarios.  In the RHS the parameter space consists of the
points $\big(M_{Z^\prime}, Re (\Delta_{R}^{bs}), Im
(\Delta_{R}^{bs}) \big)$. For a given value of $M_{Z^\prime}$ we
can use the constraints from $ \Delta M_{B_s}$ and $S_{\psi\phi}$
to show the allowed regions in the $Re (\Delta_{R}^{bs}) - Im
(\Delta_{R}^{bs})$ plane. In Fig.(\ref{ZBs}) left, we plot the
allowed regions in the $Re (\Delta_{R}^{bs}) - Im
(\Delta_{R}^{bs})$ plane for a value of $M_{Z^\prime}=1$ TeV. The
red (blue) color corresponds to the allowed regions from the
bounds on $ \Delta M_{B_s}$ ($S_{\psi\phi}$). Clearly from the
figure combining both constraints reduces the allowed regions in
the $Re (\Delta_{R}^{bs}) - Im (\Delta_{R}^{bs})$ plane.

  We turn now to the LRS. The parameter space in this case consists
 of the points the $\big(M_{Z^\prime}, Re (\Delta_{L}^{bs}), Im (\Delta_{L}^{bs})
\big),Re (\Delta_{R}^{bs}), Im (\Delta_{R}^{bs}) \big)$. For a
given value of $M_{Z^\prime}$ we can use the constraints from $
\Delta M_{B_s}$ and $S_{\psi\phi}$ to find the allowed regions in
the $\big(Re (\Delta_{L}^{bs}), Im (\Delta_{L}^{bs}),Re
(\Delta_{R}^{bs}), Im (\Delta_{R}^{bs}) \big)$ space. In
Fig.(\ref{ZBs}) middle we show the allowed regions in the $ Re
(\Delta_{R}^{bs})- Re (\Delta_{L}^{bs})$ from the bounds on $
\Delta M_{B_s}$ at $M_{Z^\prime}=1$ TeV corresponding to case $Im
(\Delta_{L}^{bs})=Im (\Delta_{R}^{bs})=0$. Their is no bound from
$S_{\psi\phi}$ in this case. In the same figure right we show the
allowed regions in the $ Im (\Delta_{R}^{bs})- Im
(\Delta_{L}^{bs})$ plane in LRS from the bounds on $ \Delta
M_{B_s}$ corresponding to case $Re(\Delta_{L}^{bs})=Re
(\Delta_{R}^{bs})=0$. Regarding the bounds from $S_{\psi\phi}$,
for this case, we find that they are so loose. This can be
explained as $Z'$ contribution to $ \Delta M_{B_s}$ is dominated
by the new LR operators which become real in this case and hence
the phase $ \theta^{B_s}_S \simeq 0$. In addition to the  previous
two cases in the LRS there is a general case where non of the real
or imaginary parts of $\Delta_{L}^{bs}$ and $\Delta_{R}^{bs}$ is
equal to zero. In Table \ref{benchmark} we list some sample sets
that satisfy both $ \Delta M_{B_s}$ and $S_{\psi\phi}$ constraints
at $M_{Z'} = 1$ TeV corresponding to this general case. In
obtaining these sets we run each of the real and imaginary parts
of $\Delta^{sb}_L$ and $\Delta^{sb}_R$ over the interval
$[-0.01,0.01]$ and require both $ \Delta M_{B_s}$ and
$S_{\psi\phi}$ constraints to be satisfied. Having discussed the
parameter space we proceed to estimate the predictions for the
$Z'$ Wilson coefficients and accordingly the branching ratios.

  We see from Eq.(\ref{c22}) that the Wilson coefficients
$C^{Z^{\prime}}_{7}$ and $\tilde{C}^{Z^{\prime}}_9$ depend on the
difference $\Delta_R^{uu}-\Delta_R^{dd}$. Clearly
$C^{Z^{\prime}}_{7}$ and $\tilde{C}^{Z^{\prime}}_9$ will vanish if
the couplings $\zeta_{R}^{qq}$, and hence  $\Delta_{R}^{qq}$, are
universal. On the other hand the maximum values of the Wilson
coefficients $C^{Z^{\prime}}_{7}$ and $\tilde{C}^{Z^{\prime}}_9$
correspond to the maximum value of the coupling difference
$\Delta_R^{uu}-\Delta_R^{dd}$.  In our analysis we assume that the
difference $\Delta_R^{uu}-\Delta_R^{dd}$ is real and
$\Delta_R^{uu}-\Delta_R^{dd} = 1$ to get an estimation of the
upper values of the branching ratios of $\bar{B}_s \rightarrow
\pi^0(\rho^0 )\,\eta^{(')} $.

 In the $RHS$ scenario scenario $\Delta^{sb}_L = 0$ and
$\Delta^{sb}_R\neq 0$. As a result $C^{Z^{\prime}}_{7} =0$ and
$\tilde{C}^{Z^{\prime}}_9 \neq 0$. This means that $Z^{\prime}$
contributes to the amplitude of the given decay process only
through the non-vanishing $\tilde{C}^{Z^{\prime}}_9$.

\begin{table}
\begin{center}
\begin{tabular}{|c|c|c|c|c|}
  \hline
  Set & $Re (\Delta_{L}^{bs})$ & $ Im (\Delta_{L}^{bs})$ &$ Re (\Delta_{R}^{bs})$ & $ Im (\Delta_{L}^{bs})$ \\
  \hline
  I &-0.01  & -0.01 &  -0.001 &  -0.001 \\
  II  &-0.01 &  0.005 &  -0.001 &  0.0005     \\
  III & 0.002 &  0.005 &  0.0005 &  0.0005  \\
  IV  &0.0035  & 0.002 &  0.0065  & -0.0055    \\
  V &0.005 &  0.005  & 0.0005 &  0.0005  \\
  \hline
\end{tabular}
\caption{Sample sets of the parameter space at $M_{Z^\prime}=1$
TeV that satisfy both $ \Delta M_{B_s}$ and $S_{\psi\phi}$
constraints. }\label{benchmark}
\end{center}
\end{table}

\begin{figure}[tbhp]
\includegraphics[width=7cm,height=7cm]{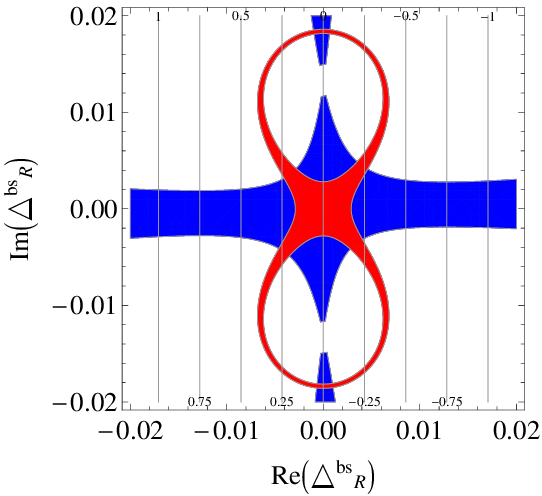}
\hspace{1.cm}
\includegraphics*[width=7cm,height=7cm]{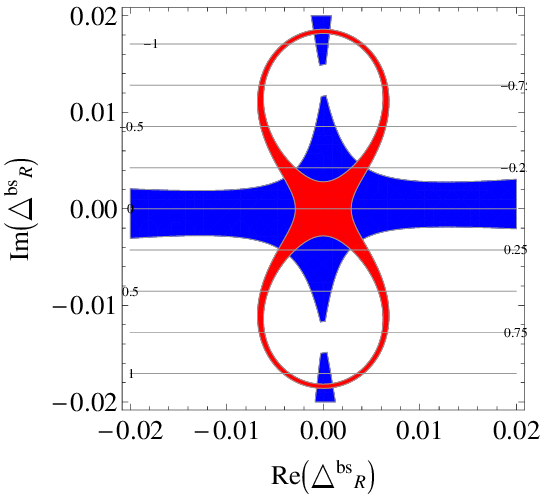}
\medskip
\caption{ Left (right): contours of the real (imaginary) part of
$\tilde{C}^{Z^{\prime}}_9$ normalized by the SM Wilson coefficient
${C}^{SM}_9$ in the RHS. The shaded red (blue) region is allowed
from the bounds on $ \Delta M_{B_s}$ ($S_{\psi\phi}$) for
$M_{Z^\prime}=1$ TeV.} \label{C9}
\end{figure}

The amplitudes of $\bar{B}_s \rightarrow \pi^0 \,\eta^{(')} $
including $Z^{\prime}$ contributions then become \bea {\cal
A}_1(\bar{B}^0_s\to \eta\, \pi^0)\times 10^6&\simeq& - \big(5.6
-1.8\, Re(\frac{\tilde{C}^{Z^{\prime}}_9}{C^{SM}_9}) -1.8\,
Im(\frac{\tilde{C}^{Z^{\prime}}_9}{C^{SM}_9})\,I\big)
C^{SM}_9\,\lambda^s_{c} -(2.3\, C^{SM}_1 + 3.7\,
C^{SM}_2 )\lambda^s_{u}\nonumber\\
{\cal A}_2(\bar{B}^0_s\to \eta\, \pi^0)\times 10^6&\simeq& -
\big(5.1-3.3\,Re(\frac{\tilde{C}^{Z^{\prime}}_9}{C^{SM}_9})
-3.3\,Im(\frac{\tilde{C}^{Z^{\prime}}_9}{C^{SM}_9})\,I\big)C^{SM}_9\,\lambda^s_{c}
- (1.6\, C^{SM}_1 + 3.4\,
C^{SM}_2)\lambda^s_{u}\nonumber\\
{\cal A}_1(\bar{B}^0_s\to \eta' \, \pi^0)\times 10^6&\simeq& \big(
0.4- 0.3\,Re(\frac{\tilde{C}^{Z^{\prime}}_9}{C^{SM}_9}) -
0.3\,Im(\frac{\tilde{C}^{Z^{\prime}}_9}{C^{SM}_9}) I
\big)C^{SM}_9\,\lambda^s_{c}
+ (0.1\, C^{SM}_1+0.3\, C^{SM}_2)\lambda^s_{u}\nonumber\\
{\cal A}_2(\bar{B}^0_s\to \eta' \, \pi^0)\times 10^6&\simeq&
\big(1.8 +6.6\,Re(\frac{\tilde{C}^{Z^{\prime}}_9}{C^{SM}_9})+6.6\,
Im(\frac{\tilde{C}^{Z^{\prime}}_9}{C^{SM}_9})I\big)C^{SM}_9\,\lambda^s_{c}
+ (2.7\, C^{SM}_1+1.2\, C^{SM}_2
)\lambda^s_{u}\nonumber\\\label{ampZ1}\eea

while for $\bar{B}_s \rightarrow \rho^0 \,\eta^{(')} $ decays we
obtain
 \bea {\cal A}_1(\bar{B}_s\to \eta\,\rho^0 )\times
10^6&\simeq& -\big(6.3- 0.3
Re(\frac{\tilde{C}^{Z^{\prime}}_9}{C^{SM}_9}) - 0.3
Im(\frac{\tilde{C}^{Z^{\prime}}_9}{C^{SM}_9})I\big) C^{SM}_9
\,\lambda^s_{c} - (3.4\, C^{SM}_1
+4.2\,C^{SM}_2 )\lambda^s_{u}\nonumber\\
{\cal A}_2(\bar{B}_s\to \eta\,\rho^0 )\times 10^6&\simeq& -\big (
3.0 -0.1 Re(\frac{\tilde{C}^{Z^{\prime}}_9}{C^{SM}_9})-0.1
Im(\frac{\tilde{C}^{Z^{\prime}}_9}{C^{SM}_9})I\big) C^{SM}_9
\,\lambda^s_{c}  - (1.6\, C^{SM}_1+2.0\, C^{SM}_2
)\lambda^s_{u}\nonumber\\{\cal A}_1(\bar{B}_s\to \eta'\,\rho^0
)\times 10^6&\simeq& \big(3.3 - 3.1
Re(\frac{\tilde{C}^{Z^{\prime}}_9}{C^{SM}_9}) - 3.1
Im(\frac{\tilde{C}^{Z^{\prime}}_9}{C^{SM}_9})I \big) C^{SM}_9
\,\lambda^s_{c} + (0.8\, C^{SM}_1+2.2\, C^{SM}_2
)\lambda^s_{u}\nonumber\\{\cal A}_2(\bar{B}_s\to \eta'\,\rho^0
)\times 10^6&\simeq& \big( 8.2 \,+5.2
Re(\frac{\tilde{C}^{Z^{\prime}}_9}{C^{SM}_9})+5.2
Im(\frac{\tilde{C}^{Z^{\prime}}_9}{C^{SM}_9})I\big)
C^{SM}_9\,\lambda^s_{c} + (6.3\, C^{SM}_1+5.4\, C^{SM}_2
)\lambda^s_{u} \label{smrho}\nonumber\\\label{ampZ2}\eea

\begin{figure}[tbhp]
\includegraphics[width=7cm,height=7cm]{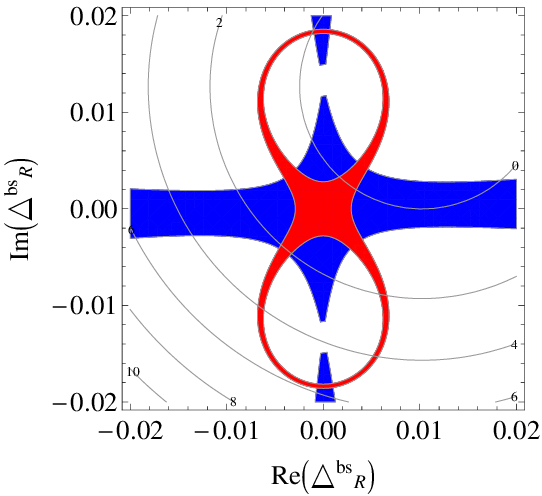}
\hspace{1.cm}
\includegraphics*[width=7cm,height=7cm]{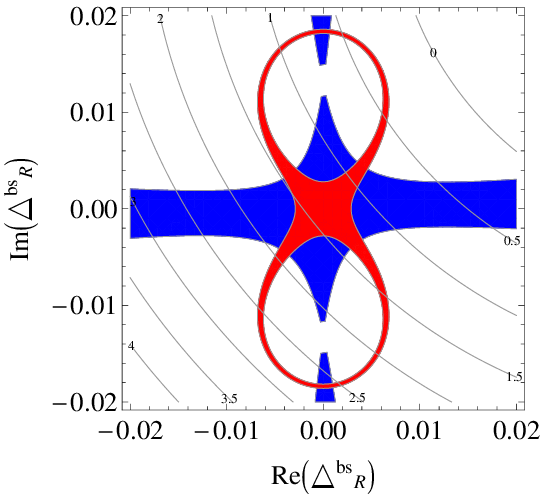}
\medskip
\caption{ Left (right): contours of ${\mathcal
R}^{\,\pi^0\,\eta'}_{2}({\mathcal R}^{\,\rho^0\,\eta'}_{2}) $ in
the RHS. The shaded red (blue) region is allowed from the bounds
on $ \Delta M_{B_s}$ ($S_{\psi\phi}$) for $M_{Z^\prime}=1$ TeV.
\label{Zam11}}
\end{figure}

We discuss now the predictions of $\tilde{C}^{Z^{\prime}}_9$. In
Fig.(\ref{C9}) we show the contours of the real and imaginary
parts of $\tilde{C}^{Z^{\prime}}_9$ normalized by the SM Wilson
coefficient ${C}^{SM}_9$. The shaded red (blue) region is allowed
from the bounds on $ \Delta M_{B_s}$ ($S_{\psi\phi}$) for a value
of $M_{Z^\prime}=1$ TeV. As can be seen from the figure, the real
part of $\tilde{C}^{Z^{\prime}}_9$ can reach a maximum value of
about $25\%$ of the SM Wilson coefficient ${C}^{SM}_9$. On the
other hand the imaginary part of $\tilde{C}^{Z^{\prime}}_9$ can
reach a maximum value equal to ${C}^{SM}_9$ at the point $\big(Re
(\Delta_{R}^{bs}) = 0,Im (\Delta_{R}^{bs})= \pm 0.018 \big)$ in
the same figure. At this point we find that $S_{\psi\phi}\simeq
0.035$ satisfying the $S_{\psi\phi}$ bound in Eq.(\ref{Sphi}).
Moreover, at the same point, we find that $[\Delta S(B_s)]_{VRR}=
-4.4\simeq -2\, S_0(x_t)$ i.e. $S_{Z'} (B_s) \simeq -2 S_{SM}
(B_s)$. Thus $\mid S_{SM} (B_s) + S_{Z'} (B_s)\mid \simeq \mid -
S_{SM} (B_s)\mid $ and thus the point $\big(Re (\Delta_{R}^{bs}) =
0,Im (\Delta_{R}^{bs}) = \pm 0.018\big )$ satisfies $ \Delta
M_{B_s}$ constraint.

As can be seen from Eqs.(\ref{ampZ1}, \ref{ampZ2}) the decay
amplitudes ${\cal A}_2(\bar{B}^0_s\to \eta' \, \pi^0)$ and ${\cal
A}_2(\bar{B}_s\to \eta'\,\rho^0 )$ have the largest coefficients
of the real and imaginary parts of $\tilde{C}^{Z^{\prime}}_9$
compared to the other amplitudes. Thus we expect that these
amplitudes receive the largest enhancements due to
$\tilde{C}^{Z^{\prime}}_9$ and consequently their branching
ratios.  We define the ratio ${\mathcal
R}^{\,M_1M_2}_{i}=\big(BR^{SM+Z'}_{i}(\bar{B}_s\to
M_1M_2)-BR^{SM}_{i}(\bar{B}_s\to
M_1M_2)\big)/BR^{SM}_{i}(\bar{B}_s\to M_1M_2)$ where $i= 1,2$
refers to solutions $1,2$ for the SCET parameter space  and $ BR$
refers to the branching ratio. The numerical value of ${\mathcal
R}^{\,M_1M_2}_{i}$ gives an estimation of the size of the
enhancement or the suppression in the branching ratios due to the
contributions of $Z'$ to the amplitude of the given decay process.
In Fig.(\ref{Zam11}) left (right) we show the contours of
${\mathcal R}^{\,\pi^0\,\eta'}_{2}({\mathcal
R}^{\,\rho^0\,\eta'}_{2}) $ over the allowed regions in the $Re
(\Delta_{R}^{bs}) - Im (\Delta_{R}^{bs})$ plane, satisfying both
$S_{\psi\phi}$ and $ \Delta M_{B_s}$ constraints, for a value of
$M_{Z^\prime}=1$ TeV. We see from Fig.(\ref{Zam11}) left that
${\mathcal R}^{\,\pi^0\,\eta'}_{2}$ can reach a maximum value of
about $6$ at the point $\big(Re (\Delta_{R}^{bs}) = 0,Im
(\Delta_{R}^{bs}) = - 0.018\big )$. This means that at this point
$Z^\prime$ contributions can enhance the total branching ratio of
$\bar B^0_s\to \,\pi^0\,\eta'$ to six times the SM prediction.
Recall that at this point $S_{\psi\phi}\simeq 0.035$ satisfying
the $S_{\psi\phi}$ bound in Eq.(\ref{Sphi}) and $S_{Z'} (B_s)
\simeq -2 S_{SM} (B_s)$ resulting in $\mid S_{SM} (B_s) + S_{Z'}
(B_s)\mid \simeq \mid - S_{SM} (B_s)\mid $ and thus $ \Delta
M_{B_s}$ constraint is  also satisfied. On the other hand from
Fig.(\ref{Zam11}) right we see that ${\mathcal
R}^{\,\rho^0\,\eta'}_{2}$ can reach a maximum value of only about
$2.5$ also at the point $\big(Re (\Delta_{R}^{bs}) = 0,Im
(\Delta_{R}^{bs}) = - 0.018\big )$. Thus the enhancement in the
total branching ratio of the decay mode $B^0_s\to \,\rho^0\,\eta'$
is not much compared to the enhancement in decay mode $\bar
B^0_s\to \,\pi^0\,\eta'$.

 We consider now LRS scenario in which $\Delta^{sb}_L\neq 0$ and
$\Delta^{sb}_R\neq 0$. As a consequence ${C}^{Z^{\prime}}_7 \neq
0$ and $ \tilde{C}^{Z^{\prime}}_9\neq 0$ and hence the amplitudes
of $\bar{B}_s \rightarrow \pi^0 \,\eta^{(')} $ can be written as
\bea {\cal A}_1(\bar{B}^0_s\to \eta\, \pi^0)\times 10^6&\simeq&
(5.6\, C^{Z^{\prime}}_7- 5.6\, C^{SM}_9+ 1.8\,
\tilde{C}^{Z^{\prime}}_9 )\lambda^s_{c} -(2.3\, C^{SM}_1 + 3.7\,
C^{SM}_2 )\lambda^s_{u}\nonumber\\
{\cal A}_2(\bar{B}^0_s\to \eta\, \pi^0)\times 10^6&\simeq& ( 5.1\,
C^{Z^{\prime}}_7 - 5.1\, C^{SM}_9+3.3\,
\tilde{C}^{Z^{\prime}}_9)\lambda^s_{c} - (1.6\, C^{SM}_1 + 3.4\,
C^{SM}_2)\lambda^s_{u}\nonumber\\{\cal A}_1(\bar{B}^0_s\to \eta'
\, \pi^0)\times 10^6&\simeq& (-0.4\, C^{Z^{\prime}}_7+ 0.4\,
C^{SM}_9- 0.3 \,\tilde{C}^{Z^{\prime}}_9)\lambda^s_{c}
+ (0.1\, C^{SM}_1+0.3\, C^{SM}_2)\lambda^s_{u}\nonumber\\
{\cal A}_2(\bar{B}^0_s\to \eta' \, \pi^0)\times 10^6&\simeq& (
-1.8 \, C^{Z^{\prime}}_7 +1.8\, C^{SM}_9+ 6.6\,
\tilde{C}^{Z^{\prime}}_9)\lambda^s_{c} + (2.7\, C^{SM}_1+1.2
C^{SM}_2 )\lambda^s_{u}\label{ampZ11}\eea while for $\bar{B}_s
\rightarrow \rho^0 \,\eta^{(')} $ decays we obtain \bea {\cal
A}_1(\bar{B}_s\to \eta\,\rho^0 )\times 10^6&\simeq& (-11.4
\,C^{Z^{\prime}}_7- 6.3\, C^{SM}_9 + 0.3\,
\tilde{C}^{Z^{\prime}}_9 )\lambda^s_{c} - (3.4\, C^{SM}_1
+4.2\,C^{SM}_2 )\lambda^s_{u} \nonumber\\
{\cal A}_2(\bar{B}_s\to \eta\,\rho^0 )\times 10^6&\simeq& (-13.3\,
C^{Z^{\prime}}_7- 3.0\, C^{SM}_9 +0.1\,
\tilde{C}^{Z^{\prime}}_9)\lambda^s_{c} - (1.6\, C^{SM}_1+2.0\,
C^{SM}_2 )\lambda^s_{u}\nonumber\\{\cal A}_1(\bar{B}_s\to
\eta'\,\rho^0 )\times 10^6&\simeq& (-1.9\,
C^{Z^{\prime}}_7+3.3\,C^{SM}_9-3.1\,\tilde{C}^{Z^{\prime}}_9
)\lambda^s_{c} + (0.8\, C^{SM}_1+2.2\, C^{SM}_2
)\lambda^s_{u}\nonumber\\{\cal A}_2(\bar{B}_s\to \eta'\,\rho^0
)\times 10^6&\simeq& (-2.6 \, C^{Z^{\prime}}_7+8.2\, C^{SM}_9
+5.2\, \tilde{C}^{Z^{\prime}}_9)\lambda^s_{c} + (6.3\,
C^{SM}_1+5.4\, C^{SM}_2 )\lambda^s_{u}\label{ampZ22}\eea

\begin{figure}[tbhp]
\includegraphics[width=7cm,height=7cm]{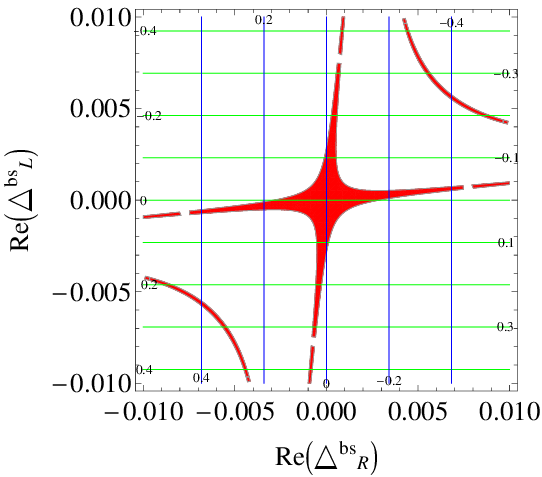}
\hspace{1.cm}
\includegraphics*[width=7cm,height=7cm]{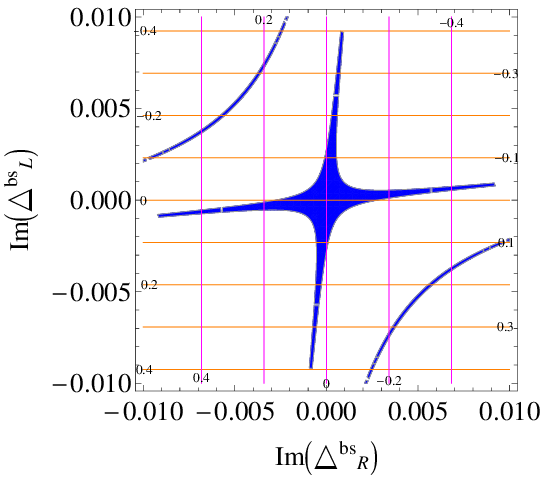}
\medskip
\caption{ Left: contours of $Re({C}^{Z^{\prime}}_7)$ ($Re(\tilde
{C}^{Z^{\prime}}_9)$) in green (blue) color normalized by the SM
Wilson coefficient ${C}^{SM}_9$ in the LRS corresponding to the
case $Im(\Delta^{sb}_R)= Im(\Delta^{sb}_L) =0 $. Right: contours
of $Im({C}^{Z^{\prime}}_7)$ ($Im(\tilde {C}^{Z^{\prime}}_9)$) in
orange (magenta) color normalized by the SM Wilson coefficient
${C}^{SM}_9$ in the LRS corresponding to the case
$Re(\Delta^{sb}_R)= Re(\Delta^{sb}_L) =0 $. In both plots the
shaded colored regions are allowed from the bounds on $ \Delta
M_{B_s}$ for  $M_{Z^\prime}=1$ TeV and .} \label{C7}
\end{figure}

We discuss now the predictions of ${C}^{Z^{\prime}}_7$ and $\tilde
{C}^{Z^{\prime}}_9$ in the LRS. We consider a case for which
$Im(\Delta^{sb}_R)= Im(\Delta^{sb}_L) =0 $. In this case $
{C}^{Z^{\prime}}_7\equiv Re({C}^{Z^{\prime}}_7)$ and  $\tilde
{C}^{Z^{\prime}}_9 \equiv Re(\tilde {C}^{Z^{\prime}}_9)$.  In
Fig.(\ref{C7}) left we show the contours of
$Re({C}^{Z^{\prime}}_7)$ ($Re(\tilde {C}^{Z^{\prime}}_9)$) in
green (blue) color normalized by the SM Wilson coefficient
${C}^{SM}_9$. The shaded red regions are allowed from the bounds
on $ \Delta M_{B_s}$ for $M_{Z^\prime}=1$ TeV as discussed before.
The contours of $Re({C}^{Z^{\prime}}_7)$ ($Re(\tilde
{C}^{Z^{\prime}}_9)$) are straight lines as ${C}^{Z^{\prime}}_7$
($\tilde {C}^{Z^{\prime}}_9$) is a function of $Re(\Delta^{sb}_L)$
($Re(\Delta^{sb}_R)$) only. Clearly from the figure
$Re({C}^{Z^{\prime}}_7)$ can reach a  maximum value around $ 0.4\,
{C}^{SM}_9$ while $Re(\tilde {C}^{Z^{\prime}}_9)$ can reach a
maximum value around $0.6\, {C}^{SM}_9$. However
$\big(Re({C}^{Z^{\prime}}_7), Re(\tilde {C}^{Z^{\prime}}_9)\big) =
\big( 0.4, 0.6\big)$ are excluded by the bounds on $ \Delta
M_{B_s}$ that require $Re(\Delta^{sb}_R)$ and $
Re(\Delta^{sb}_L)$, and hence $Re({C}^{Z^{\prime}}_7)$ and
$Re(\tilde {C}^{Z^{\prime}}_9)$, not to be large simultaneously.

We consider another case where $Re(\Delta^{sb}_R)=
Re(\Delta^{sb}_L) =0 $. In this case $ {C}^{Z^{\prime}}_7\equiv
Im({C}^{Z^{\prime}}_7)$ and  $\tilde {C}^{Z^{\prime}}_9 \equiv
Im(\tilde {C}^{Z^{\prime}}_9)$. In Fig.(\ref{C7}) right we show
the contours of $Im({C}^{Z^{\prime}}_7)$ ($Im(\tilde
{C}^{Z^{\prime}}_9)$) in orange (magenta) color normalized by the
SM Wilson coefficient ${C}^{SM}_9$. The shaded blue regions are
allowed from the bounds on $ \Delta M_{B_s}$ for $M_{Z^\prime}=1$
TeV. Recall that the constraints from $S_{\psi\phi}$ are so loose
as discussed before. The conclusion for this case is the same as
the previous case with just doing the replacements
$Re({C}^{Z^{\prime}}_7)\rightarrow Im({C}^{Z^{\prime}}_7)$ and
$Re(\tilde {C}^{Z^{\prime}}_9)\rightarrow Im(\tilde
{C}^{Z^{\prime}}_9)$.

 We finally consider the general case where none of the real or
imaginary parts of $\Delta^{sb}_L$ and $\Delta^{sb}_R$ is equal to
zero. In Table \ref{benchmarkWil} we list the predictions of
${C}^{Z^{\prime}}_7$ and $\tilde{C}^{Z^{\prime}}_9$ corresponding
to some sample sets of the parameter space allowed by both $\Delta
M_{B_s}$ and $S_{\psi\phi}$ constraints for $M_{Z'}= 1$ TeV. As
before, in obtaining these results we run each of the real and
imaginary parts of $\Delta^{sb}_L$ and $\Delta^{sb}_R$ over the
interval $[-0.01,0.01]$ requiring that both $ \Delta M_{B_s}$ and
$S_{\psi\phi}$ constraints to be satisfied. From the Table we note
that $Re(\tilde {C}^{Z^{\prime}}_9)= Im(\tilde {C}^{Z^{\prime}}_9)
\simeq 0.6\,C^{SM}_9$ corresponding to set III of the allowed
parameter space. This is the maximum value of $Re(\tilde
{C}^{Z^{\prime}}_9)\big(Im(\tilde {C}^{Z^{\prime}}_9)\big)$
obtained in our scan for all points in the parameter space that
satisfy $ \Delta M_{B_s}$ and $S_{\psi\phi}$ constraints.
Regarding ${C}^{Z^{\prime}}_7$ we find that
$Re({C}^{Z^{\prime}}_7)\big(Im({C}^{Z^{\prime}}_7)\big)$ can reach
a maximum value around $ 0.4\, {C}^{SM}_9$.

\begin{table}
\begin{center}
\begin{tabular}{|c|c|c|c|c|c|c|c|c|}
  \hline
  Set & $Re (\Delta_{L}^{bs})$ & $ Im (\Delta_{L}^{bs})$ &$ Re (\Delta_{R}^{bs})$ & $ Im (\Delta_{L}^{bs})$
  & $Re({C}^{Z^{\prime}}_7)$& $Im({C}^{Z^{\prime}}_7) $& $Re(\tilde{C}^{Z^{\prime}}_9)$& $Im(\tilde{C}^{Z^{\prime}}_9)$ \\
  \hline
  I & -0.01 &  -0.01 &  -0.001 &  -0.001 &  0.432943 &  0.432943  &   0.0586095 &  0.0586095   \\
  II  & -0.01  & 0.005 &  -0.001 &  0.0005 &  0.432943 &  -0.216471  &   0.0586095 &  -0.0293047    \\
  III & -0.001 &  -0.001 &  -0.01 &  -0.01  & 0.0432943 &  0.0432943 &    0.586095 &  0.586095   \\
  IV  & 0.0035  & -0.001 &  0.0095 &  0.005 &  -0.15153  & 0.0432943  &   -0.55679 &  -0.293047   \\
  V & 0.0005  & -0.001  & 0.005  & -0.01 &  -0.0216471 &  0.0432943  &   -0.293047 &  0.586095   \\
  \hline
\end{tabular}
\caption{Predictions of ${C}^{Z^{\prime}}_7$ and
$\tilde{C}^{Z^{\prime}}_9$ corresponding to some sample sets of
the parameter space for $M_{Z^\prime}=1$ TeV allowed by both
$\Delta M_{B_s}$ and $S_{\psi\phi}$ constraints.
}\label{benchmarkWil}
\end{center}
\end{table}

\begin{figure}[tbhp]
\includegraphics[width=7cm,height=7cm]{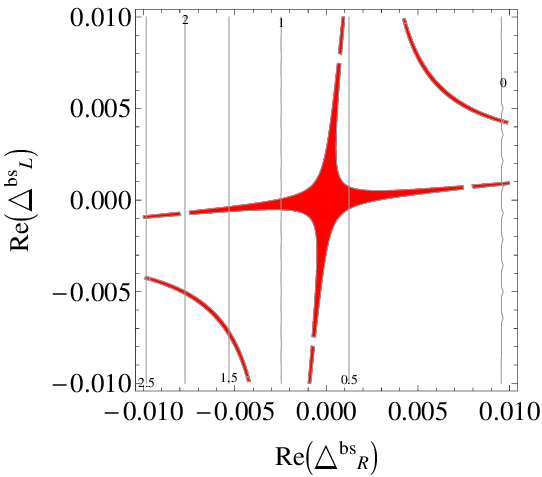}
\hspace{1.cm}
\includegraphics*[width=7cm,height=7cm]{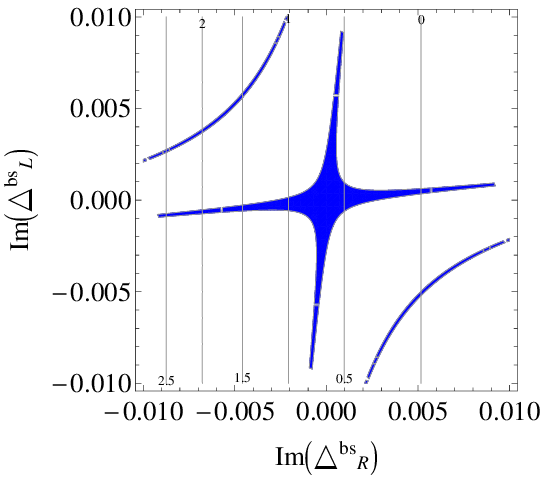}
\medskip
\caption{ Left: contours of ${\mathcal R}^{\,\pi^0\,\eta'}_{2}$ in
the LRS for the case $Im(\Delta^{sb}_L)=Im(\Delta^{sb}_R)=0$.
Right: contours of ${\mathcal R}^{\,\pi^0\,\eta'}_{2}$ in the LRS
for the case $Re(\Delta^{sb}_L)=Re(\Delta^{sb}_R)=0$. In both
plots the shaded colored regions satisfy the bounds on $ \Delta
M_{B_s}$ for $M_{Z^\prime}=1$ TeV. \label{ZLRSp}}
\end{figure}

Finally we turn to the predictions of the branching ratios of the
processes under consideration.  We start with $\bar{B}^0_s\to
\eta(\eta')\, \pi^0$ decays. Their related amplitudes are given in
Eq.(\ref{ampZ11}). Clearly the amplitude ${\cal
A}_2(\bar{B}^0_s\to \eta'\, \pi^0)$ has the largest coefficient of
$\tilde{C}^{Z^{\prime}}_9$ compared to the coefficients of both
the ${C}^{Z^{\prime}}_7$ and ${C}^{SM}_9$ in the other amplitudes.
As a result this amplitude receives the largest enhancement due to
$Z^{\prime}$ contributions. In Fig.(\ref{ZLRSp}) left, we show the
contours of ${\mathcal R}^{\,\pi^0\,\eta'}_{2}$ in the LRS for the
case $Im(\Delta^{sb}_L)=Im(\Delta^{sb}_R)=0$ where the shaded red
regions satisfy the bounds on $ \Delta M_{B_s}$. In the same
figure right, we show the contours of ${\mathcal
R}^{\,\pi^0\,\eta'}_{2}$ for the case
$Re(\Delta^{sb}_L)=Re(\Delta^{sb}_R)=0$ where the shaded blue
regions are allowed by the bounds on $ \Delta M_{B_s}$. In both
plots we take $M_{Z^\prime}=1$ TeV. We see from the figure that
${\mathcal R}^{\,\pi^0\,\eta'}_{2} $ can reach a maximum value of
about $2.5$ in both cases.

In Table \ref{Branshpi} we list the predictions of ${\mathcal
R}^{\,\pi^0\,\eta'}_{2}$ corresponding to the general case where
non of the real or imaginary parts of $\Delta_{L}^{bs}$ and
$\Delta_{R}^{bs}$ is equal to zero and are allowed by both $
\Delta M_{B_s}$ and $S_{\psi\phi}$ constraints. As before, in
obtaining these results we run each of the real and imaginary
parts of $\Delta^{sb}_L$ and $\Delta^{sb}_R$ over the interval
$[-0.01,0.01]$ requiring that both $ \Delta M_{B_s}$ and
$S_{\psi\phi}$ constraints to be satisfied for $M_{Z^\prime}=1$
TeV. From the Table we note that ${\mathcal
R}^{\,\pi^0\,\eta'}_{2}\simeq 4.6$ corresponding to set I of the
allowed parameter space. This means that $Z^\prime$ contributions
can enhance the total branching ratio of $B^0_s\to\,\pi^0\,\eta'$
$4.6$ times the SM prediction. This is the maximum value we
obtained in our scan for all sets of the parameter space that
satisfy both $ \Delta M_{B_s}$ and $S_{\psi\phi}$ constraints.

\begin{figure}[tbhp]
\includegraphics[width=7cm,height=7cm]{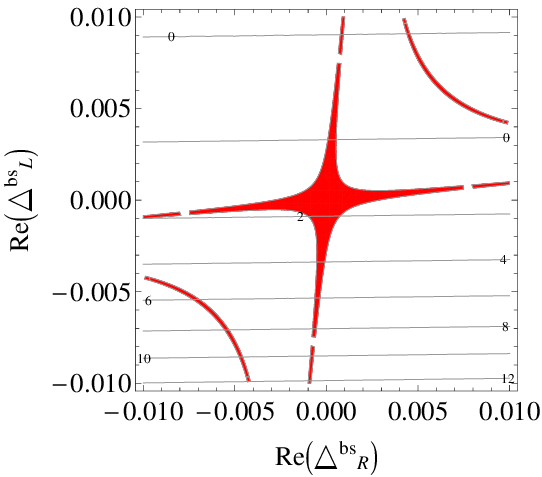}
\hspace{1.cm}
\includegraphics*[width=7cm,height=7cm]{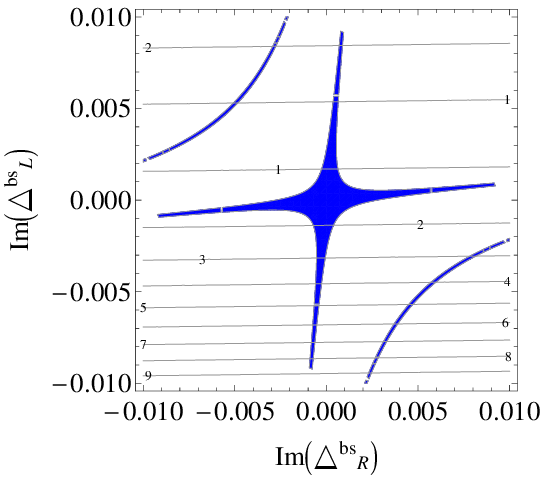}
\medskip
\caption{ Left: contours of ${\mathcal R}^{\,\eta\,\rho^0}_{2}$ in
the LRS for the case $Im(\Delta^{sb}_L)=Im(\Delta^{sb}_R)=0$.
Right: contours of ${\mathcal R}^{\,\rho^0\,\eta}_{2} $ in the LRS
for the case $Re(\Delta^{sb}_L)=Re(\Delta^{sb}_R)=0$. In both
plots the shaded colored regions satisfy the bounds on $ \Delta
M_{B_s}$ for $M_{Z^\prime}=1$ TeV.\label{ZLRSr}}
\end{figure}

We turn now to the decay modes $\bar{B}^0_s\to \eta(\eta')\,
\rho^0$. Their  related amplitudes are given in Eq.(\ref{ampZ22}).
We note that the amplitude ${\cal A}_2(\bar{B}^0_s\to \eta\,
\rho^0)$ has the largest coefficient of ${C}^{Z^{\prime}}_7$
compared to the coefficients of both the
$\tilde{C}^{Z^{\prime}}_9$ and ${C}^{SM}_9$ in the other
amplitudes. As a result this amplitude receives the largest
enhancement due to $Z^{\prime}$ contributions. In
Fig.(\ref{ZLRSr}) left, we show the contours of ${\mathcal
R}^{\,\rho^0\,\eta}_{2}$ in the LRS for the case
$Im(\Delta^{sb}_L)=Im(\Delta^{sb}_R)=0$. In the same figure right,
we show the contours of ${\mathcal R}^{\,\rho^0\,\eta}_{2}$ in the
LRS for the case $Re(\Delta^{sb}_L)=Re(\Delta^{sb}_R)=0$. In the
figure the shaded colored regions satisfy the bounds on $ \Delta
M_{B_s}$ for $M_{Z^\prime}=1$ TeV. We see from the figure that
$Z^\prime$ contributions can enhance the total branching ratio of
$B^0_s\to\,\rho^0\,\eta$ by about one order of magnitude comparing
to the SM prediction. In fact this is the conclusion also for the
general case where non of the real or imaginary parts of
$\Delta_{L}^{bs}$ and $\Delta_{R}^{bs}$ is equal to zero as shown
in Table \ref{Branshr}. In that Table we list the predictions of
${\mathcal R}^{\,\rho^0\,\eta}_{2}$ corresponding to some sample
sets of the parameter space  allowed by both $ \Delta M_{B_s}$ and
$S_{\psi\phi}$ constraints. As before, in obtaining these results
we run each of the real and imaginary parts of $\Delta^{sb}_L$ and
$\Delta^{sb}_R$ over the interval $[-0.01,0.01]$ requiring that
both $ \Delta M_{B_s}$ and $S_{\psi\phi}$ constraints are
satisfied at a value $M_{Z^\prime}=1$ TeV. From the Table we see
that $Z^\prime$ contributions can enhance the total branching
ratio of $B^0_s\to\,\rho^0\,\eta$ up to one order of magnitude
comparing to the SM prediction.

\begin{table}
\begin{center}
\begin{tabular}{|c|c|c|c|c|c|}
\hline Set & $Re (\Delta_{L}^{bs})$ & $ Im (\Delta_{L}^{bs})$ &$
Re (\Delta_{R}^{bs})$ & $ Im (\Delta_{L}^{bs})$ & ${\mathcal
R}^{\,\pi^0\,\eta'}_{2}$ \\
\hline
I & -0.001& -0.001& -0.01&   -0.01&   4.6  \\
II  &-0.001&   0.002&   -0.0025& -0.01& 3.3   \\
III & 0.0005&   -0.001&   0.002&   -0.01&   2.6   \\
IV & -0.004&   0.0005&   -0.01&   -0.0025&   2.8\\
V & -0.001&   0.0035&   -0.001&   -0.007&   2.4\\
 \hline
\end{tabular}
\caption{ Predictions for ${\mathcal R}^{\,\pi^0\,\eta'}_{2}$
corresponding to some sample sets of the parameter space for
$M_{Z^\prime}=1$ TeV allowed by both $\Delta M_{B_s}$ and
$S_{\psi\phi}$ constraints. }\label{Branshpi}
\end{center}
\end{table}

Finally, from Fig.(\ref{ZLRSr}) we note that the enhancement of
the branching ratios by an order of magnitude occurs in the
regions in the parameter space corresponding to the thin branches
in the figure. In these regions the $[\Delta S(B_s)]_{VLL(VRR)}$
and $[\Delta S(B_s)]_{LR}$ contributions to $B_s-\bar B_s$ mixing
given in Eqs. (\ref{VLVR},\ref{VLR}) cancel each other to a large
extent. For a proper interpretation of the given upper limit for
the enhancement it is thus necessary to know the degree of
fine-tuning between $\Delta^{sb}_L$ and $\Delta^{sb}_R$ for the
corresponding points in the parameter space. For the case of real
$\Delta^{sb}_L$ and $\Delta^{sb}_R$, the fine-tuning can be
quantified by the measure $X_{B_s}$ introduced in eq.(26) in
Ref.\cite{Crivellin:2015era}. The corresponding expression reads

\bea X_{B_s}&=& \frac{(\Delta^{sb}_L)^2+(\Delta^{sb}_R)^2 -
b_{B_s}\Delta^{sb}_L\Delta^{sb}_R}{(\Delta^{sb}_L)^2+(\Delta^{sb}_R)^2
+ b_{B_s} \Delta^{sb}_L\Delta^{sb}_R}\label{Xbse}\eea

where

\be b_{B_s} = \frac{(\lambda^s_t)^2 g^2_{SM}}{4\tilde{r}
T(B_s)}\big[C^{LR}_1(\mu_{Z'})\langle
Q^{LR}_1(\mu_{Z'},B_s)\rangle+C^{LR}_2(\mu_{Z'})\langle
Q^{LR}_2(\mu_{Z'},B_s)\rangle\big] \ee

\begin{figure}[tbhp]
\begin{center}
\includegraphics*[width=7cm,height=7cm]{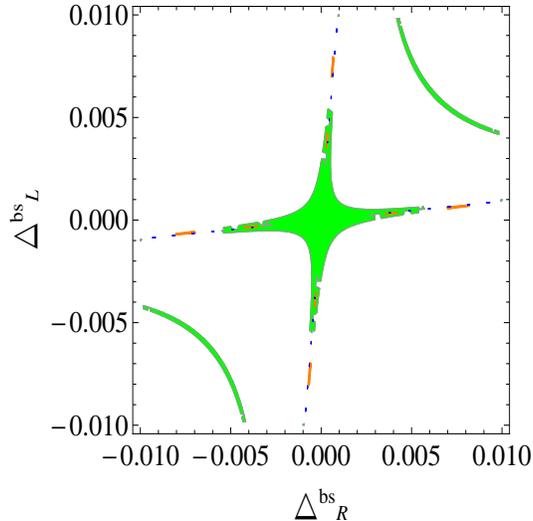}
\medskip
\caption{  Points satisfying $X_{B_s} < 10$, $10 < X_{B_s} < 100$
and $100 < X_{B_s}$ in green, orange and blue colors respectively.
All colored points satisfy the bounds on $ \Delta M_{B_s}$ for
$M_{Z^\prime}=1$ TeV.\label{XBs12p}}
\end{center}
\end{figure}

At $ M_{Z^\prime} = 1$ TeV we find that $b_{B_s} \simeq -10.8$. In
Fig.(\ref{XBs12p}) we show the points in the
$\Delta^{sb}_L-\Delta^{sb}_R$ satisfying $X_{B_s} < 10$, $10 <
X_{B_s} < 100$ and $100 < X_{B_s}$ in green, orange and blue
colors respectively. All colored points satisfy the bounds on $
\Delta M_{B_s}$ for $M_{Z^\prime}=1$ TeV. Clearly from
Figs.(\ref{ZLRSr},\ref{XBs12p}) if we exclude points that lead to
$100 < X_{B_s}$ from the parameter space and exclude their
corresponding predictions of the branching ratios we still can
have an enhancement by an order of magnitude.  As before this
enhancement occurs for  rather fine-tuned points where $ \Delta
M_{B_s}$ constraint on $\mid S_{SM} (B_s) + S_{Z'} (B_s)\mid $ is
fulfilled by overcompensating the SM via $S_{Z'} (B_s) \simeq -2
S_{SM} (B_s)$. For instances this enhancement is still allowed for
the the points lies on the thin red curved branch in
Fig.(\ref{ZLRSr}) for which $\Delta^{sb}_R < -0.04$ and
$\Delta^{sb}_L < -0.06$. In fact this conclusion agrees with the
findings of Ref.\cite{Crivellin:2015era} where they found that
sizeable enhancements in the branching ratios of the process under
their consideration are still possible  for a fine-tuning of
$X_{B_s} \leq  100$.

\section{Conclusion \label{sec:conclusion}}

In this work we have studied the decay modes $\bar{B}_s
\rightarrow \pi^0(\rho^0 )\,\eta^{(')} $  within a model with an
additional $U(1)'$~ gauge symmetry and adopting  SCET as a
framework to calculate the amplitudes. We have derived the
contributions to the amplitudes, within Soft Collinear Effective
Theory, arising from new physics contributions to the weak
effective Hamiltonian.

 In the study we have considered a leptophobic $Z'$ boson where its couplings to
leptons vanish. Such a model can appear in models with an $E_6$
gauge symmetry. In this model $Z'$ mass is much less constrained
and the strongest constraints on the parameter space can be
obtained by considering $B_s-\overline{B}_s$ mixing. We considered
two scenarios, the Right-Handed Scenario, RHS, and Left-Right
Scenario, LRS, based on the couplings of $Z'$ to quarks that
appear in the Wilson coefficients. In these scenarios we discussed
the constraints on the parameter space from considering
$B_s-\overline{B}_s$ mixing. As a consequence, we presented the
predictions of the Wilson coefficients and accordingly the
branching ratios of $\bar{B}_s \rightarrow \pi^0(\rho^0
)\,\eta^{(')} $.

 In the RHS we found that the real part of
$\tilde{C}^{Z^{\prime}}_9$ can reach a maximum value of about
$25\%$ of ${C}^{SM}_9$. On the other hand the imaginary part of
$\tilde{C}^{Z^{\prime}}_9$ can reach a maximum value equal to
${C}^{SM}_9$. As a result we found that the decay amplitudes
${\cal A}_2(\bar{B}^0_s\to \pi^0 \eta')$ and ${\cal
A}_2(\bar{B}_s\to \rho^0\,\eta' )$ receive the largest
enhancements due to the contributions of
$\tilde{C}^{Z^{\prime}}_9$ as they have the largest coefficients
of the real and imaginary parts of $\tilde{C}^{Z^{\prime}}_9$
compared to the other amplitudes. Accordingly we found that
$Z^\prime$ contributions can enhance the total branching ratio of
$\bar B^0_s\to \,\pi^0\,\eta'$ to six times the SM prediction
while for  $B^0_s\to \,\rho^0\,\eta'$ it is just $2.5$ times the
SM prediction. This kind of enhancement occurs for a rather
fine-tuned point where $ \Delta M_{B_s}$ constraint on $\mid
S_{SM} (B_s) + S_{Z'} (B_s)\mid $ is fulfilled by overcompensating
the SM via $S_{Z'} (B_s) \simeq -2 S_{SM} (B_s)$. Moreover the
constraint from $S_{\psi\phi}$ is also satisfied as $Z^\prime$
coupling to the $b$ and $s$ quarks is real at this point.

\begin{table}
\begin{center}
\begin{tabular}{|c|c|c|c|c|c|}
\hline Set & $Re (\Delta_{L}^{bs})$ & $ Im (\Delta_{L}^{bs})$ &$
Re (\Delta_{R}^{bs})$ & $ Im (\Delta_{L}^{bs})$ & ${\mathcal
R}^{\,\rho^0\,\eta}_{2}$ \\
\hline
I & -0.01&   0.002&   -0.001&   0.0005&   11.7  \\
II  &-0.01& -0.0025& -0.004&   0.0005&     13.3   \\
III & -0.01&   0.0035& -0.001& 0.0005& 11.6  \\
IV & -0.0085&   0.0035&   -0.004&   -0.001&   9.3\\
V& 0.002&   -0.01&   0.0005&   -0.001  & 8.7  \\
VI & -0.0025&   -0.01&   -0.001&   0.002&   11.5\\
\hline
\end{tabular}
\caption{ Predictions for ${\mathcal R}^{\,\rho^0\,\eta}_{2}$
corresponding to some sample sets of the parameter space for
$M_{Z^\prime}=1$ TeV allowed by both $\Delta M_{B_s}$ and
$S_{\psi\phi}$ constraints. }\label{Branshr}
\end{center}
\end{table}
In the LRS the parameter space consists of the real and imaginary
parts of $\Delta^{sb}_L$ and $\Delta^{sb}_R$ in addition to
$M_{Z'}$.  At a value of $M_{Z'}= 1$ TeV  we scanned the real and
imaginary parts of $\Delta^{sb}_L$ and $\Delta^{sb}_R$ over the
interval $[-0.01,0.01]$ requiring that both $ \Delta M_{B_s}$ and
$S_{\psi\phi}$ constraints to be satisfied. As a consequence we
found that the maximum enhancements in the $Z'$ Wilson
coefficients correspond to $Re(\tilde {C}^{Z^{\prime}}_9)=
Im(\tilde {C}^{Z^{\prime}}_9) \simeq 0.6\,C^{SM}_9$ and $Re(
{C}^{Z^{\prime}}_7)= Im({C}^{Z^{\prime}}_7) \simeq 0.4\,C^{SM}_9$.
Regarding the branching ratios we found that $Z^\prime$
contributions can enhance the branching ratio of
$B^0_s\to\,\pi^0\,\eta'$ by about $4.6$ times the SM prediction.
Moreover we found that the branching ratio of
$B^0_s\to\,\rho^0\,\eta$ can be enhanced up to one order of
magnitude comparing to the SM prediction for several sets of the
parameter space satisfying both $ \Delta M_{B_s}$ and
$S_{\psi\phi}$ constraints and for a fine-tuning of $X_{B_s} \leq
100$.  For these points $ \Delta M_{B_s}$ constraint on $\mid
S_{SM} (B_s) + S_{Z'} (B_s)\mid $ is fulfilled by overcompensating
the SM via $S_{Z'} (B_s) \simeq -2 S_{SM} (B_s)$.

\newpage
\section*{Acknowledgements} This work is  supported by
the research grant NTU-ERP-102R7701. I would like to thank prof.
H.~-Y.~Cheng for useful discussions and his suggestion to do this
study.


\begin{thebibliography}{99}

\bibitem{Fleischer:1994rs}
  R.~Fleischer,
  Phys.\ Lett.\ B {\bf 332}, 419 (1994).
\bibitem{Deshpande:1994yd}
  N.~G.~Deshpande, X.~-G.~He and J.~Trampetic,
  Phys.\ Lett.\ B {\bf 345}, 547 (1995)
  [hep-ph/9410388].


\bibitem{Chen:1998dta}
  Y.~-H.~Chen, H.~-Y.~Cheng and B.~Tseng,
  Phys.\ Rev.\ D {\bf 59}, 074003 (1999)
  [hep-ph/9809364].

\bibitem{Cheng:2009mu}
  H.~-Y.~Cheng and C.~-K.~Chua,
  Phys.\ Rev.\ D {\bf 80}, 114026 (2009)
  [arXiv:0910.5237 [hep-ph]].



\bibitem{Beneke:2003zv}
  M.~Beneke and M.~Neubert,
  Nucl.\ Phys.\ B {\bf 675}, 333 (2003)
  [hep-ph/0308039].




\bibitem{Hofer:2010ee}
  L.~Hofer, D.~Scherer and L.~Vernazza,
  JHEP {\bf 1102}, 080 (2011)
  [arXiv:1011.6319 [hep-ph]].

\bibitem{Ali:2007ff}
  A.~Ali, G.~Kramer, Y.~Li, C.~D.~Lu, Y.~L.~Shen, W.~Wang and Y.~M.~Wang,
  Phys.\ Rev.\  D {\bf 76}, 074018 (2007)
  [arXiv:hep-ph/0703162].


 \bibitem{Wang:2008rk}
  W.~Wang, Y.~M.~Wang, D.~S.~Yang and C.~D.~Lu,
  Phys.\ Rev.\  D {\bf 78}, 034011 (2008)
  [arXiv:0801.3123 [hep-ph]].

\bibitem{Faisel:2011kq}
  G.~Faisel,
  JHEP {\bf 1208}, 031 (2012)
  [arXiv:1106.4651 [hep-ph]].


\bibitem{Faisel:2013nra}
  G.~Faisel,
  Phys.\ Lett.\ B {\bf 731}, 279 (2014)
  [arXiv:1311.0740 [hep-ph]].







\bibitem{Descotes-Genon:2013wba}
  S.~Descotes-Genon, J.~Matias and J.~Virto,
  Phys.\ Rev.\ D {\bf 88}, 074002 (2013)
  doi:10.1103/PhysRevD.88.074002
  [arXiv:1307.5683 [hep-ph]].


\bibitem{Gauld:2013qba}
  R.~Gauld, F.~Goertz and U.~Haisch,
  Phys.\ Rev.\ D {\bf 89}, 015005 (2014)
  doi:10.1103/PhysRevD.89.015005
  [arXiv:1308.1959 [hep-ph]].

\bibitem{Buras:2013qja}
  A.~J.~Buras and J.~Girrbach,
  JHEP {\bf 1312}, 009 (2013)
  [arXiv:1309.2466 [hep-ph]].


\bibitem{Gauld:2013qja}
  R.~Gauld, F.~Goertz and U.~Haisch,
  JHEP {\bf 1401}, 069 (2014)
  doi:10.1007/JHEP01(2014)069
  [arXiv:1310.1082 [hep-ph]].


\bibitem{Buras:2013dea}
  A.~J.~Buras, F.~De Fazio and J.~Girrbach,
  JHEP {\bf 1402}, 112 (2014)
  doi:10.1007/JHEP02(2014)112
  [arXiv:1311.6729 [hep-ph]].

\bibitem{Altmannshofer:2014cfa}
  W.~Altmannshofer, S.~Gori, M.~Pospelov and I.~Yavin,
  Phys.\ Rev.\ D {\bf 89}, 095033 (2014)
  doi:10.1103/PhysRevD.89.095033
  [arXiv:1403.1269 [hep-ph]].
















\bibitem{Xiao:2006gf}
  Z.~-j.~Xiao, X.~Liu and H.~-s.~Wang,
  Phys.\ Rev.\ D {\bf 75}, 034017 (2007)
  [hep-ph/0606177].





\bibitem{Sun:2002rn}
  J.~-f.~Sun, G.~-h.~Zhu and D.~-s.~Du,
  Phys.\ Rev.\ D {\bf 68}, 054003 (2003)
  [hep-ph/0211154].



\bibitem{Williamson:2006hb}
  A.~R.~Williamson and J.~Zupan,
  Phys.\ Rev.\  D {\bf 74}, 014003 (2006)
  [Erratum-ibid.\  D {\bf 74}, 03901 (2006)]
  [arXiv:hep-ph/0601214].










\bibitem{Zhang:2000ic}
  D.~Zhang, Z.~-j.~Xiao and C.~S.~Li,
  Phys.\ Rev.\ D {\bf 64}, 014014 (2001)
  [hep-ph/0012063].



\bibitem{Bauer:2000ew}
  C.~W.~Bauer, S.~Fleming and M.~E.~Luke,
  Phys.\ Rev.\  D {\bf 63}, 014006 (2000)
  [arXiv:hep-ph/0005275].


\bibitem{Bauer:2000yr}
  C.~W.~Bauer, S.~Fleming, D.~Pirjol and I.~W.~Stewart,
  Phys.\ Rev.\  D {\bf 63}, 114020 (2001)
  [arXiv:hep-ph/0011336].



  \bibitem{Chay:2003zp}
  J.~Chay and C.~Kim,
  Phys.\ Rev.\ D {\bf 68}, 071502 (2003)
  [arXiv:hep-ph/0301055].

\bibitem{Chay:2003ju}
  J.~Chay and C.~Kim,
  Nucl.\ Phys.\ B {\bf 680}, 302 (2004)  [arXiv:hep-ph/0301262].


\bibitem{Bauer:2005kd}
  C.~W.~Bauer, I.~Z.~Rothstein and I.~W.~Stewart,
  Phys.\ Rev.\  D {\bf 74}, 034010 (2006)
  [arXiv:hep-ph/0510241].



\bibitem{Jain:2007dy}
  A.~Jain, I.~Z.~Rothstein and I.~W.~Stewart,
  arXiv:0706.3399 [hep-ph].





\bibitem{Michael:2013gka}
  C.~Michael {\it et al.} [ETM Collaboration],
  Phys.\ Rev.\ Lett.\  {\bf 111}, no. 18, 181602 (2013)
  doi:10.1103/PhysRevLett.111.181602
  [arXiv:1310.1207 [hep-lat]].


\bibitem{Ball:2007rt}
  P.~Ball and G.~W.~Jones,
  JHEP {\bf 0703}, 069 (2007)
  doi:10.1088/1126-6708/2007/03/069
  [hep-ph/0702100 [HEP-PH]].

\bibitem{Bakulev:2003cs}
  A.~P.~Bakulev, S.~V.~Mikhailov and N.~G.~Stefanis,
  Phys.\ Lett.\ B {\bf 578}, 91 (2004)
  doi:10.1016/j.physletb.2003.10.033
  [hep-ph/0303039].











\bibitem{Masip:1999mk}
  M.~Masip and A.~Pomarol,
  Phys.\ Rev.\ D {\bf 60}, 096005 (1999)
  [hep-ph/9902467].

\bibitem{E6}
 E.~Nardi, Phys.\ Rev.\ D {\bf 48} (1993) 1240 [hep-ph/9209223];
 J.~Bernabeu, E.~Nardi and D.~Tommasini, Nucl.\ Phys.\ B  {\bf409} (1993) 69
 [hep-ph/9306251];
 V.~D.~Barger, M.~S.~Berger and R.~J.~Phillips, Phys.\ Rev.\ D {\bf52} (1995) 1663
 [hep-ph/9503204];
 M.~B.~Popovic and E.~H.~Simmons, Phys.\ Rev.\ D {\bf62} (2000) 035002
 [hep-ph/0001302];
 T.~G.~Rizzo Phys.\ Rev.\ D {\bf59} (1999) 015020 [hep-ph/9806397].

\bibitem{Langacker:2000ju}
  P.~Langacker and M.~Plumacher,
  Phys.\ Rev.\ D {\bf 62}, 013006 (2000)
  [hep-ph/0001204].

\bibitem{Chaudhuri:1994cd}
  S.~Chaudhuri, S.~-W.~Chung, G.~Hockney and J.~D.~Lykken,
  Nucl.\ Phys.\ B {\bf 456}, 89 (1995)
  [hep-ph/9501361].

\bibitem{Cleaver:1997jb}
  G.~Cleaver, M.~Cvetic, J.~R.~Espinosa, L.~L.~Everett and P.~Langacker,
  Nucl.\ Phys.\ B {\bf 525}, 3 (1998)
  [hep-th/9711178].

\bibitem{Cleaver:1998gc}
  G.~Cleaver, M.~Cvetic, J.~R.~Espinosa, L.~L.~Everett, P.~Langacker and J.~Wang,
  Phys.\ Rev.\ D {\bf 59}, 055005 (1999)
  [hep-ph/9807479].


\bibitem{Chang:2013hba}
  Q.~Chang, X.~Q.~Li and Y.~D.~Yang,
  J.\ Phys.\ G {\bf 41}, 105002 (2014)
  [arXiv:1312.1302 [hep-ph]].





\bibitem{Altmannshofer:2009ma}
  W.~Altmannshofer, A.~J.~Buras, D.~M.~Straub and M.~Wick,
  JHEP {\bf 0904}, 022 (2009)
  [arXiv:0902.0160 [hep-ph]].



\bibitem{Buras:2012jb}
  A.~J.~Buras, F.~De Fazio and J.~Girrbach,
  JHEP {\bf 1302}, 116 (2013)
  [arXiv:1211.1896 [hep-ph]].



\bibitem{Buras:2012dp}
  A.~J.~Buras, F.~De Fazio, J.~Girrbach and M.~V.~Carlucci,
  JHEP {\bf 1302}, 023 (2013)
  [arXiv:1211.1237 [hep-ph]].




\bibitem{Buras:2013rqa}
  A.~J.~Buras, F.~De Fazio, J.~Girrbach, R.~Knegjens and M.~Nagai,
  JHEP {\bf 1306}, 111 (2013)
  [arXiv:1303.3723 [hep-ph]].

\bibitem{Buras:2013td}
  A.~J.~Buras, J.~Girrbach and R.~Ziegler,
  JHEP {\bf 1304}, 168 (2013)
  [arXiv:1301.5498 [hep-ph]].


\bibitem{Buras:2013uqa}
  A.~J.~Buras, R.~Fleischer, J.~Girrbach and R.~Knegjens,
  JHEP {\bf 1307}, 77 (2013)
  [arXiv:1303.3820 [hep-ph]].



\bibitem{Buras:2014fpa}
  A.~J.~Buras, J.~Girrbach-Noe, C.~Niehoff and D.~M.~Straub,
  JHEP {\bf 1502}, 184 (2015)
  [arXiv:1409.4557 [hep-ph]].







\bibitem{Langacker:2008yv}
  P.~Langacker,
  Rev.\ Mod.\ Phys.\  {\bf 81}, 1199 (2009)
  [arXiv:0801.1345 [hep-ph]].



\bibitem{Grossman:1999av}
  Y.~Grossman, M.~Neubert and A.~L.~Kagan,
  JHEP {\bf 9910}, 029 (1999)
  [hep-ph/9909297].

\bibitem{Langacker1}
 P.~Langacher and M.~Pl\"{u}macher, Phys.\ Rev.\ D {\bf 62} (2000) 013006 [hep-ph/0001204].



\bibitem{Barger:2009eq}
  V.~Barger, L.~Everett, J.~Jiang, P.~Langacker, T.~Liu and C.~Wagner,
  Phys.\ Rev.\ D {\bf 80}, 055008 (2009)
  [arXiv:0902.4507 [hep-ph]].


\bibitem{Chang:2009wt}
  Q.~Chang, X.~Q.~Li and Y.~D.~Yang,
  JHEP {\bf 0905}, 056 (2009)
  [arXiv:0903.0275 [hep-ph]].

\bibitem{Barger:2009qs}
  V.~Barger, L.~L.~Everett, J.~Jiang, P.~Langacker, T.~Liu and C.~E.~M.~Wagner,
  JHEP {\bf 0912}, 048 (2009)
  [arXiv:0906.3745 [hep-ph]].

\bibitem{Langacker2}
 V.~Barger, L.~Everett, J.~Jiang, P.~Langacker, T.~Liu and C.~Wagner, JHEP {\bf 0912} (2009) 048 [arXiv:0906.3745].




 \bibitem{chiang1}
 C.~W.~Chiang, Y.~F.~Lin, Jusak Tandean, JHEP {\bf 1111} (2011) 083 [arXiv:1108.3969].





\bibitem{Rizzo:1998ut}
  T.~G.~Rizzo,
  Phys.\ Rev.\ D {\bf 59}, 015020 (1998)
  [hep-ph/9806397].






\bibitem{Buras:2001ra}
  A.~J.~Buras, S.~Jager and J.~Urban,
  Nucl.\ Phys.\ B {\bf 605}, 600 (2001)
  [hep-ph/0102316].

\bibitem{Buras:2012fs}
  A.~J.~Buras and J.~Girrbach,
  JHEP {\bf 1203}, 052 (2012)
  [arXiv:1201.1302 [hep-ph]].


\bibitem{Amhis:2012bh}
  Y.~Amhis {\it et al.} [Heavy Flavor Averaging Group Collaboration],
  arXiv:1207.1158 [hep-ex].


\bibitem{Chatrchyan:2012oaa}
  S.~Chatrchyan {\it et al.} [CMS Collaboration],
  Phys.\ Lett.\ B {\bf 720}, 63 (2013)
  [arXiv:1212.6175 [hep-ex]].



\bibitem{Clarke:2012hhi}
P. Clarke, Results on CP-violation in Bs Mixing,
LHCb-TALK-2012-029, CERN, Geneva Switzerland (2012).





\bibitem{Crivellin:2015era}
  A.~Crivellin, L.~Hofer, J.~Matias, U.~Nierste, S.~Pokorski and J.~Rosiek,
  Phys.\ Rev.\ D {\bf 92}, no. 5, 054013 (2015)
  doi:10.1103/PhysRevD.92.054013
  [arXiv:1504.07928 [hep-ph]].





\end{thebibliography}
\end{document}